\documentclass{article}
\usepackage[utf8]{inputenc} 
\usepackage[T1]{fontenc}    
\usepackage[english]{babel}
\usepackage{csquotes}
\usepackage{arxiv}
\usepackage{hyperref}       
\usepackage{url}            
\usepackage{booktabs}       
\usepackage{amsfonts}       
\usepackage{xfrac}       
\usepackage{microtype}      
\usepackage{graphicx}

\usepackage[style=alphabetic, backend=biber]{biblatex}

\usepackage{amsmath}
\usepackage{amssymb}
\usepackage{doi}
\usepackage{subcaption}
\usepackage{xcolor}
\usepackage{bm}
\usepackage[title]{appendix}
\usepackage{nicefrac}
\usepackage{physics}

\allowdisplaybreaks 

\definecolor{linkcol}{HTML}{024d6e}
\definecolor{citeherecol}{HTML}{026e4c}
\definecolor{warncol}{HTML}{e32909}
\definecolor{emph}{HTML}{c22004}
\definecolor{comment_color}{HTML}{ff306c}

\hypersetup{
  colorlinks=True,
  linkcolor=linkcol,
  citecolor=linkcol,
urlcolor=linkcol}

\AtEveryBibitem{%
  \clearlist{language}%
  \clearfield{url}%
  \clearfield{isbn}%
  \clearfield{issn}%
  \clearfield{note}%
  \clearfield{howpublished}%
  \clearfield{urldate}%
  \clearfield{extra}%
  \clearfield{eprinttype}%
  \clearfield{eprintclass}%
  \clearfield{eprint}%
}
\DeclareFieldFormat{urldate}{}

\newcommand{\diff}[1]{\mathrm{d}#1\!\mathop{}}
\newcommand{\difff}[1]{\mathcal{D}#1\!\mathop{}}
\newcommand{\mcom}{\ ,}
\newcommand{\mdot}{\ .}

\newcommand{\avg}[1]{\left< #1 \right>}

\renewcommand{\v}[1]{\bm{#1}}
\newcommand{\obs}{\mathcal{O}}
\newcommand{\act}{\mathcal{A}}
\newcommand{\Dt}{\Delta t}
\newcommand{\Dx}{\Delta x}

\newcommand{\nicehalf}{\nicefrac{1}{2}}
\renewcommand{\epsilon}{\varepsilon}

\title{Entropy Production from Density Field Theories for interacting particles systems}
\author{Antonin Brossollet, Giulio Biroli}

\begin{document}
\maketitle
\begin{abstract}
Entropy production quantifies the breaking of time-reversal symmetry in non-equilibrium systems. Here, we develop a direct method to obtain closed, tractable expressions for entropy production in a broad class of dynamical density functional theories—from Dean's exact stochastic equation for microscopic densities to coarse-grained fluctuating-hydrodynamics models with density-dependent mobility. The method employs an Onsager-Machlup path-integral formulation. Our results reproduce particle-level calculations and matches recent Doi-Peliti treatments, confirming that the irregular noise structure of Dean's equation poses no obstacle when handled consistently. We further extend the framework to active mixtures with non-reciprocal interactions and to run-and-tumble or active-Brownian suspensions, generalizations that require a careful treatment of the spurious-drift. Our method furnishes a practical route to quantify irreversibility in density functional field theories and paves the way for systematic studies of entropy production in multi-field active fluids that couple density, momentum and orientation. 
\end{abstract}

\section{Introduction}

Over the last twenty-five years a sequence of theoretical breakthroughs has created a new chapter of non-equilibrium statistical physics known as \emph{stochastic thermodynamics}.  The discovery of the Jarzynski equality and Crooks fluctuation relation for small driven systems \cite{jarzynskiNonequilibriumEqualityFree1997,crooksEntropyProductionFluctuation1999} demonstrated that entropy production could be related to experimentally accessible work statistics far from equilibrium.  These results were soon followed up by Lebowitz and Spohn's large-deviation treatment of stochastic dynamics \cite{lebowitzGallavottiCohenType1999} and by Seifert's formulation of trajectory-level first- and second-law statements for Markov processes \cite{seifertStochasticThermodynamicsFluctuation2012a}.  Subsequent work generalised the theory to coarse-grained systems \cite{espositoEntropyProductionCorrelation2010}, information-to-work conversion \cite{parrondoThermodynamicsInformation2015}, and coarse-grained field theories \cite{nardiniEntropyProductionField2017}.  These developments have fueled up the interest in \emph{entropy production} as a central quantity that diagnoses time-reversal symmetry breaking and sets bounds on the non-equilibrium nature of irreversible processes.

A natural context where entropy production plays a central role is active matter. In fact, active matter—collections of self-propelled particles or agents that convert stored or ambient energy into systematic motion—provides a paradigmatic arena in which non-equilibrium steady states (NESS) arise at macroscopic scales \cite{marchettiHydrodynamicsSoftActive2013,catesMotilityInducedPhaseSeparation2015}.  Because active particles continuously dissipate energy, their coarse-grained dynamics violate detailed balance even in steady state; quantifying this violation has become a key objective.  Pioneering studies used stochastic thermodynamics to estimate entropy production from single-particle orientations and currents \cite{fodorHowFarEquilibrium2016}; field-theoretic approaches soon followed, revealing spatial structure and scale-dependent irreversibility in scalar active models \cite{nardiniEntropyProductionField2017,maggiCriticalActiveDynamics2022}.  More recent work has employed Doi-Peliti formalism \cite{pruessnerFieldTheoriesActive2022} to access entropy production in microscopic particle systems. 

Here we develop an exact framework for computing entropy production directly for \emph{density field theories} that describe interacting particles. Our targets are both exact density field theories \`a la Dean \cite{deanLangevinEquationDensity1996a} and coarse-grained density functional theories -- a widespread framework to study fluids out of equilibrium \cite{tevrugtClassicalDynamicalDensity2020}.
We start our analysis focusing on Dean's stochastic partial differential equation for the microscopic density of interacting Langevin particles \cite{deanLangevinEquationDensity1996a}. Within this setting, we obtain a simple expression for the steady-state entropy production rate, which for particles interacting via two-body potentials can be written in terms of one- and two-point density correlation functions, in agreement with the Doi-Peliti treatment of \cite{pruessnerFieldTheoriesActive2022}.  We then generalise the method to  dynamical density functional theories and to fluctuating-hydrodynamics equations with density-dependent mobility.  We verify consistency with (i) direct particle-level calculations, and (ii) recent Doi-Peliti analyses \cite{pruessnerFieldTheoriesActive2022}.  Our work demonstrates that the irregular nature of Dean's field theory does not hinder the computation of entropy production, and it opens the way to systematic studies of irreversibility in density functional field theories containing several coarse-grained fields, such as e.g. density, momentum and orientation.

\section{Stochastic equations on the density field and Onsager-Machlup path integral formulation}
The aim of this work is to obtain a general expression for the entropy production for dynamical field theories associated to interacting particles. 
We focus on the framework of Dynamical Density Functional Theory (DDFT), which models the time evolution of the one-body density field $\rho(\mathbf{x}, t)$ \cite{tevrugtClassicalDynamicalDensity2020}. DDFT reduces the complex dynamics of many-body systems to the evolution of a single scalar field. DDFT encompasses a variety of formulations: some are phenomenological \cite{evansNatureLiquidvapourInterface1979,fraaijeDynamicDensityFunctional1993,kirkpatrickConnectionsKineticEquilibrium1987}, others arise from coarse-graining procedures such as fluctuating hydrodynamics \cite{kawasakiMicroscopicAnalysesDynamical1998,moriTransportCollectiveMotion1965}, and yet others provide an exact reformulation of the underlying particle dynamics, as in Dean's equation \cite{deanLangevinEquationDensity1996a, archerDynamicalDensityFunctional2004}. Depending on the context, DDFT focuses on the full stochastic one-body density or its ensemble-averaged counterpart. The former leads to a stochastic partial differential equation (SPDE), while the latter results in a deterministic PDE.

We begin our analysis by considering the case in which DDFT serves as an exact reformulation of the stochastic Langevin dynamics of  $N$  interacting point-like particles interacting by a two-body potential $V(\v{x},\v{y})=V(\v{x}-\v{y})$. The corresponding evolution equation for the density field  $\rho$ defined as
\begin{equation}\label{eq:density_delta}
  \rho(\v{x},t) = \sum_{i=1}^{N} \delta(\v{x} - \v{X}_i(t)) \mcom
\end{equation}
was first derived by Dean \cite{deanLangevinEquationDensity1996a}. It reads:
\begin{equation}\label{eq:dean_equilibrium}
\begin{gathered}
  \partial_t \rho(\v{x},t) = \dot{\rho}(\v{x},t) = \nabla \cdot  \left(\rho(\v{x},t) \nabla \frac{\delta \mathcal{F}}{\delta \rho(\v{x})} \right) + \nabla \cdot ( \sqrt{2T \rho(\v{x},t)} \v{\eta}(\v{x},t)) \\
  \mathcal{F}[\rho] = \frac{1}{2} \int \diff{\v{x}} \diff{\v{y}} \rho(\v{x}) V(\v{x}-\v{y}) \rho(\v{y}) + T \int \diff{\v{x}} \rho(\v{x}) \log \rho(\v{x}) \\
  \avg{\eta^{\alpha} (\v{x},t) \eta^{\beta}(\v{x'},t')} = \delta^{\alpha \beta} \delta(\v{x}-\v{x'}) \delta(t-t') \mdot
\end{gathered}
\end{equation}
Here $\mathcal{F}[\rho]$ is the (field-dependent) free energy. The field $\v{\eta}(\v{x},t)$ is white in both space and time. $T$ is the temperature of the bath, equal to the variance of the noise in equilibrium due to the Einstein relation (with the proper choice of units such that the Boltzmann constant $k_B=1$). This will still be used to designate the variance of the noise also in out-of-equilibrium situations. The stochastic equation has to be interpreted following Ito's prescription. 

Dean's equation was originally derived for systems with reversible dynamics. In what follows, we focus on the irreversible case, where particles are subject to a non-conservative force field and evolve according to the following Langevin equations:
\begin{equation}\label{eq:langevin}
\begin{gathered}
  \frac{\diff{}}{\diff{t}} \v{X}_i(t) = - \sum_{j \neq i} \nabla V(\v{X}_i(t) - \v{X}_j(t)) + \v{f}(\v{X}_i(t)) + \v{\xi}_i(t) \\
  \big<\xi_i^\alpha(t) \xi_j^\beta(t')\big> = 2T \delta_{ij} \delta^{\alpha \beta} \delta(t-t') \mdot
\end{gathered}
\end{equation}
The force field $\v{f}$ is non-conservative because it is not the gradient of a potential. Note that in order to be able to express the problem in terms of the density, the non-conservative force has to be independent of the particle index. Otherwise, each particle would have a specific dynamics and would not be indistinguishable, which in turns means a single one-body density could not encode the necessary information about the system dynamics.

From the set of Langevin equations with added non-conservative forces one can directly obtain Dean's equation in this irreversible case (see appendix): 
\begin{equation}\label{eq:dean_outeq}
  \dot{\rho} = \nabla \cdot  \left(\rho \Big(\nabla \frac{\delta \mathcal{F}}{\delta \rho} - \v{f}(\v{x}) \Big) \right) + \nabla \cdot \Big( (2T \rho)^{\nicehalf} \v{\eta} \Big) \mdot
\end{equation}
Written as is, where the density is defined as a sum of delta peaks \eqref{eq:density_delta}, this equation is exact, albeit mathematically difficult to handle due to its very irregular nature \cite{lehmannDeanKawasakiDynamicsIllposedness2018}. This should be seen as a formal expression and as such can be formally manipulated. 
In cases where DDFT arises from a coarse-grained description, the resulting equations are well-defined and regular. The theoretical framework we develop for Dean's equation can therefore be naturally extended to these smoother, coarse-grained formulations—as we will discuss later.

A crucial step to obtain an expression for the entropy production rate is to derive an Onsager-Machlup (OM) path-integral formulation of the stochastic equation~\eqref{eq:dean_outeq}. The computation is pretty standard, but for the sake of clarity it is sketched in App. \ref{app:OM_action_proof}. It follows the usual route: we express the average of a generic observable of the density, imposing via a Dirac delta function that the field configuration is a solution of the dynamical equation. Then using the fact that the noise field is Gaussian, hence that we know explicitly its probability distribution, we perform a change of variable to integrate out the noise field. It results in the expression for the average of an arbitrary observable of the density:
\begin{equation}
  \avg{\obs} =  \int \difff{\rho} \obs[\rho] \frac{1}{\mathcal{N[\rho]}}e^{-\act[\rho]} \mcom
\end{equation}
where the OM action is
\begin{equation}
  \act[\rho] = \frac{1}{4T} \int \diff{t}\diff{\v{x}} \left( \frac{1}{\rho} \nabla^{-1} \dot{\rho} - \nabla \frac{\delta \mathcal{F}}{\delta \rho} + \v{f} \right) \left( \nabla^{-1} \dot{\rho} - \rho(\nabla \frac{\delta \mathcal{F}}{\delta \rho} - \v{f}) \right) 
\end{equation}
and $\mathcal{N}[\rho]$ is a field dependent normalization factor, which exact expression depends on the specific discretization of the stochastic differential equation. As we explain in depth in the Appendix, this normalization factor will not play a role in the computation of the entropy production. In consequence, and for simplicity, we do not give its explicit expression here. 
The operator $\nabla^{-1}$ is formally the inverse of the gradient or divergence operator: for a scalar field $X$, we have $\nabla \nabla^{-1} X = X$. Said otherwise, $\nabla^{-1}$ maps the scalar field $X$ to a vector field $Y$ such that $\nabla \v{Y} = X$.\footnote{This mapping is not unique, thus a gauge condition is necessary to have a proper definition. This choice will not play any role in what follows. A natural choice would be to impose $\nabla \wedge \v{Y} = 0$.}  Equipped with the action $\act[\rho]$, the trajectory weight can be defined in the usual manner. For a path $\{\rho(\v{x},t)\}_{0 \leq t \leq \tau}$ it is 
\begin{equation}
  P[\rho] = \frac{1}{\mathcal{N}[\rho]} \exp \left( -\act[\rho] \right) \mdot
\end{equation}

 We refer to the appendix \ref{app:technical_details} for details about discretization \cite{cugliandoloRulesCalculusPath2017,lauStatedependentDiffusionThermodynamic2007}. Due to the choice of 
considering an Itô discretization for the dynamics, the expression of $\act[\rho]$ is simpler (see \ref{app:technical_details}) than in more general cases.

\section{Entropy production rate for Dean's dynamical field theory}\label{sec:entropy_DDFT}
In this section we obtain a general expression for the entropy production rate in the case of Dean's equation. 
We begin with a brief introduction to the concept of stochastic entropy production, which has attracted considerable attention in recent years—particularly in the study of active matter—as a powerful tool to quantify the local irreversibility of dynamical processes \cite{seifertStochasticThermodynamicsFluctuation2012a}. At the heart of this approach is the idea of formulating a general framework to locally measure the breaking of time-reversal symmetry (TRS). This is essential, as TRS is a defining characteristic of equilibrium systems; thus, identifying how and where this symmetry is violated in out-of-equilibrium systems is key to probe the physical mechanisms at play. 

\subsection{General introduction: forward and backward paths, and entropy production}
To probe the TRS breaking and compute the entropy production rate, the first step is to define the time-reversed trajectory $\{\rho^R(\v{x},t)\}_{0 \leq t \leq \tau}$ and obtain the associated path probability within the field theory. The time reversal transformation is defined as
\begin{equation}
  \begin{aligned}
    t &\rightarrow \tau - t \\
    \rho(\v{x},t) &\rightarrow \rho^R(\v{x},t) = \rho(\v{x}, \tau-t) \mdot
  \end{aligned}
\end{equation}
Note that when considering the time-reversal operation, we still consider the original dynamics for the reversed path. More precisely, we do not consider that some other parameters of the system change sign under this operation. We are only interested in observing the reverse of the original trajectory, under the forward dynamics. In general, for a system in which the dynamics does not respect TRS, this reversed trajectory is very atypical, i.e. with very low probability. For instance, when the noise is weak, it has to go against the deterministic forces driving the system in a specific direction. 

Considering Eq.~\eqref{eq:dean_outeq}, it is clear that the only antisymmetric part under the time-reversal transformation is $\partial_t \rho$. Hence, the path probability of the reversed path, as a functional of the forward path is easily computed to be
\begin{gather}
  \begin{gathered}
    P^R[\rho] = \frac{1}{\mathcal{N}[\rho^R]}\exp \left( -\act^R[\rho] \right) \\
    \act^R[\rho] = \frac{1}{4T} \int \diff{t}\diff{\v{x}} \left( -\frac{1}{\rho} \nabla^{-1} \dot{\rho} - \nabla \frac{\delta \mathcal{F}}{\delta \rho} + \v{f} \right) \left( - \nabla^{-1} \dot{\rho} - \rho(\nabla \frac{\delta \mathcal{F}}{\delta \rho} - \v{f}) \right) + \mathcal{A}_J[\rho] \mcom
  \end{gathered}
\end{gather}
where $\mathcal{A}_J[\rho]$ is a contribution to the action coming from the change in discretization from 
anti-Itô to Itô (the former is obtained when plugging the time-reversed trajectory, but the latter is needed is order to have a coherent and unique discretization for all expressions).  
Again, the details of this discussion are deferred to App. \ref{app:technical_details} to simplify the presentation. 

A caveat, in the expression for the dynamical action the gradients and time derivatives do not act on everything lying on their right, but simply on what is just after them. Expressions are written using simply functions and not operators.\footnote{Everything could be recast in terms of operators by performing appropriate integration by parts, but this is an unnecessary complication for the present work.}

According to the theory of stochastic thermodynamics \cite{seifertStochasticThermodynamicsFluctuation2012a, nardiniEntropyProductionField2017}, the entropy production associated to a given stochastic path $\{\rho(\v{x},t)\}_{0 \leq t \leq \tau}$ reads:  
\begin{equation}\label{eq:path_entropy}
  \hat{S}_\tau[\rho] = \log \left( \frac{P[\rho]}{P^R[\rho]} \right) = \act^R[\rho] - \act[\rho] + \log \left( \frac{\mathcal{N}[\rho^R]}{\mathcal{N}[\rho]} \right)   \mdot
\end{equation}
Note that we consider paths conditioned on the initial condition. One should also take into account the probability of the initial condition in order to have a vanishing entropy production in equilibrium. This extra-term does not give any contribution to the entropy production rate in the NESS, so we stick to the definition $\hat{S}_\tau[\rho]$ above. 
The entropy production rate between $\tau$ and $\tau+d\tau$ is then defined by considering the limit of an infinitesimally small-time 
\begin{equation}\label{eq:entropy_prod_rate}
  \dot{\hat{S}}_\tau[\rho] = \lim_{d\tau \to 0} \frac{1}{d\tau}\left(\hat{S}_{\tau+d\tau}[\rho] -\hat{S}_\tau[\rho]\right) \mdot
\end{equation}

An important point to note, as emphasized by the hat used in the above definition, is that $\hat{S}$ is a realization dependent quantity, defined at the level of a single density path. A quantity of physical interest is often the ensemble average of this path dependent quantity, as discussed for instance in \cite{seifertStochasticThermodynamicsFluctuation2012a,cocconiEntropyProductionExactly2020}. In the following we will focus on the average entropy production.

We will also restrict our analysis to Non Equilibrium Steady States (NESS), which are defined as stationary solutions of the Fokker-Planck equation associated to the SPDE \eqref{eq:dean_outeq}. These steady states have in general no reason to correspond to thermal equilibrium when the system do not respect detailed balance, as it is the case here. Therefore, they are characterized by a non-zero entropy production. Our results below can be easily generalised to cases in which the system is not in a steady state.

\subsection{Entropy production and ensemble average}
From the definition of the entropy for a specific path \eqref{eq:path_entropy}, it is easy to derive, using the explicit form of the forward and the time-reversed actions, and using spatial integration by parts, a compact expression, as detailed in the App. \ref{app:DDFT_entropy_simplification}
\begin{equation}
  \hat{S}_\tau [\rho] = -\frac{1}{T} \int_0^\tau \diff{t} \int \diff{\v{x}} \left[ \frac{\delta \mathcal{F}}{\delta \rho} \dot{\rho} + \v{f} (\nabla^{-1} \dot{\rho})\right] + \mathcal{A}_J + \log\left( \frac{\mathcal{N}[\rho^R]}{\mathcal{N}[\rho]} \right) \mdot
\end{equation}
The last term, the logarithms of the normalization factor can be shown to be equal to zero (see the Appendix), hence we will drop  it henceforth. 
The products of stochastic quantities are, as in the rest of the paper, interpreted in the Itô sense. This means that ensemble average are straightforward to compute, using the Itô prescription. To make things as explicit as possible, we perform the average in two steps. First, we average over the field increment, $\dot{\rho}$ conditioning on the value of the field at time $t$. Due to the Itô choice, $\dot{\rho}$ is normally distributed, with mean $\nabla (\rho (\nabla \frac{\delta \mathcal{F}}{\delta \rho} - \v{f}))=\nabla(\rho(\rho \ast \nabla V + T \frac{\nabla \rho}{\rho} - \v{f}))$, where $\ast$ denote the convolution product. In a second step we perform the average over the field at time $t$, using the steady-state probability measure $P_{\text{ss}}[\rho]$.

An important caveat, due to the Itô interpretation, we cannot use the standard chain rule to write time derivatives. It needs to be altered according to the well-known Itô's lemma. It turns out, as detailed in the Appendix, that for the specific case of computing the entropy production, the contributions originating from $\mathcal{A}_J$ are exactly the right terms stemming from Itô's lemma, which allow to write the first term as the time derivative of $\mathcal{F}$. This leads to 
\begin{equation}\label{eq:DDFT_entropy_production_stoch}
\begin{aligned}
  \hat{S}_\tau [\rho] &= -\frac{1}{T} \int_0^\tau \diff{t} \int \diff{\v{x}} \left[ \partial_t \mathcal{F} + \v{f} (\nabla^{-1} \dot{\rho})\right]  \\
  &= - \frac{1}{T} \int_{0}^\tau \diff{t} \int \diff{\v{x}} (\v{f} \nabla^{-1} \dot{\rho}) - \frac{1}{T} \Delta \mathcal{F} \mcom
\end{aligned}
\end{equation}
where $\Delta \mathcal{F} = \mathcal{F}(\tau) - \mathcal{F}(0)$ is the variation of the free-energy functional. Note that in absence of a non-equilibrium drive, one recover the correct detailed balance result. 

Leveraging the remark we made above, it is straightforward to take the ensemble average to compute the entropy production. In the following we consider systems in a steady state. Hence, we focus on computing the ensemble average of the entropy production rate \eqref{eq:entropy_prod_rate}. In a steady state, the average of the variation of the free-energy is trivially zero. The average of the logarithm of the ratio of the normalization factors will also not contribute: the steady-state probability distribution is invariant, hence $\big<\mathcal{N}[\rho]\big> = \big<\mathcal{N}[\rho^R]\big>$. More details are provided in App. \ref{app:normalization_factor}. Taking the $\tau \to 0$ limit, substituting $\dot{\rho}$ by its expression using Dean's equation and performing the ensemble average leads to the following expressions of the average entropy production rate
\begin{equation}\label{eq:avg_entropy_prod}
\begin{aligned}
  \dot{S} = \big< \dot{\hat{S}} \big> &= -\frac{1}{T} \int \difff{\rho} \int \diff{\v{x}} \v{f}(\v{x,t}) \rho(\v{x},t) \Big[ \nabla \frac{\delta \mathcal{F}}{\delta \rho}(\v{x},t) - \v{f}(\v{x},t)  \Big] P_{\textbf{ss}}[\rho]  \\
  & =-\frac{1}{T}  \int \difff{\rho} \int \diff{\v{x}} \v{f}(\v{x},t) \rho(\v{x},t) \Big[(\rho \ast \nabla V)(\v{x},t) + T \frac{\nabla \rho (\v{x},t)}{\rho(\v{x},t)} - \v{f}(\v{x},t)\Big] P_\text{ss}[\rho] \mcom
\end{aligned}
\end{equation}
where $P_{\text{ss}}[\rho]$ is the steady-state distribution of the density field. This can be simplified further to give an expression in terms of the one and two point correlation functions, due to the fact that particles only interact pair-wise:
\begin{equation}
  \begin{aligned}
    \dot{S} &= - \frac{1}{T} \int \diff{\v{x}} \Big[ \avg{\int \diff{\v{y}} \v{f}(\v{x}) \rho(\v{x}) \nabla V(\v{x}-\v{y}) \rho(\v{y})} + T \avg{ \v{f}(\v{x}) \nabla \rho(\v{x})} - \avg{\v{f}^2(\v{x}) \rho(\v{x})} \Big] \\
    &= - \frac{1}{T} \int \diff{\v{x}} \diff{\v{y}} \v{f}(\v{x},t) \nabla V(\v{x}-\v{y}) G_2(\v{x},\v{y}) + \int \diff{\v{x}} \nabla \v{f}(\v{x}) G_1(\v{x}) + \frac{1}{T} \int \diff{\v{x}} \v{f}^2(\v{x}) G_1(\v{x}) \mdot
  \end{aligned}
\end{equation}
$G_1(\v{x})$ and $G_2(\v{x},\v{y})$ are the one-point and two-point steady-state correlation function of the density field. 

This is the first main result of this paper. It is a compact form of the entropy production associated to the non-equilibrium Dean equation~\eqref{eq:dean_outeq}. The fact that we could write a form involving only the one and two point correlation functions crucially comes from the pair-potential we used. It is however straightforward to adapt the result to take into account more general potentials. The final expression would then depend on higher-order correlation functions of the density.

\section{Comparison with other methods}

The goal of this section is to compare our approach with the interesting framework developed in \cite{pruessnerFieldTheoriesActive2022}, which starts from a different field-theoretic formulation of the underlying Langevin dynamics—specifically, a Doi-Peliti field theory. We also demonstrate that our general result agrees with a direct computation of entropy production for a system of  $N$ particles evolving under Langevin dynamics. Since Dean's equation provides an exact description of the microscopic system, both approaches should yield consistent results—and indeed they do, thereby confirming the validity of our method, which later on we will apply to coarse-grained versions of DDFT, for which no direct connection to a particles system is available.

\subsection{Doi-Peliti field theory approach}\label{subsec:Doi-Peliti}
In \cite{pruessnerFieldTheoriesActive2022},  Garcia-Millan and Pruessner  develop a theoretical framework to compute the entropy production rate associated to a system of active particles. They start from the Fokker-Planck equation of the system (which can be unambiguously associated to a Langevin equation) and derive from it a Doi-Peliti field theory. The main advantage and motivation of the Doi-Peliti approach of \cite{pruessnerFieldTheoriesActive2022} is that it preserves by design the particle nature of the constituents of the system \cite{botheParticleEntityDoiPeliti2023}. This point is crucial, as the choice of degrees of freedom considered has a significant impact when computing a system's entropy production. If one relies on a coarse-grained description of the underlying dynamics, some small-scale irreversibility may be lost, potentially leading to inaccurate or underestimated values of the entropy production.

Coarse-grained field theory for particle systems, as the one developed in \cite{nardiniEntropyProductionField2017} can therefore only access a part of the total entropy production, as discussed in the conclusion of \cite{pruessnerFieldTheoriesActive2022}\footnote{We emphasize that there is nothing inherently problematic about using a coarse-grained field theory; however, it is important to recognize that part of the entropy production—specifically, the contribution from the microscopic degrees of freedom that have been integrated out—is no longer accessible in such a description.}.

As explicitly shown in \cite{tjhungClusterPhasesBubbly2018}, to derive the evolution equation of the Active Model B+ (AMB+) as an effective description of a particle system, a key approximation is to suppress the density dependence of the mobility, which transform the multiplicative noise into an additive one. This step leads to an evolution equation not respecting anymore the particle nature of the constituents of the model. By working with Dean equation, we avoid this simplification and as we illustrate now, our formalism allows to recover the correct expression for the entropy production obtained in \cite{pruessnerFieldTheoriesActive2022}. 

In order to make explicit the relationship with \cite{pruessnerFieldTheoriesActive2022} we start from \eqref{eq:avg_entropy_prod} and explicitly use the form the density \eqref{eq:density_delta}, by plugging the expression in the result for the entropy production. To match the setup studied in \cite{pruessnerFieldTheoriesActive2022}, we add an external potential, denoted $2\Upsilon(\v{x})$.\footnote{The factor 2 is simply due to the difference in the way we define the potential with respect to \cite{pruessnerFieldTheoriesActive2022}, in terms of the summation over the indices.} This leads to (we drop some $t$ dependencies for ease of notation):
\begin{multline}
  S_\tau = -\frac{1}{T} \int \ \prod_i \diff{\v{X}_i} \int \diff{t} \diff{\v{x}} \v{f}(\v{x}) \Big[ \sum_i \delta(\v{x}-\v{X}_i(t)) \int \diff{\v{y}} \sum_j \delta(\v{y}-\v{X}_j(t)) \nabla V(\v{x}-\v{y}) \\
  + 2 \nabla \Upsilon(\v{x}) \sum_i \delta(\v{x}-\v{X}_i(t)) + T \sum_i \nabla \delta(\v{x}-\v{X}_i(t)) - \v{f}(\v{x}) \sum_i \delta(\v{x}-\v{X}_i(t)) \Big] P_{\text{ss}}(\{\v{X}_i\}) \mdot
\end{multline}
Note that here $P_{\text{ss}}$ is strictly speaking not the same distribution as the one appearing in \eqref{eq:avg_entropy_prod}, because it is express in terms of $\v{X}_i$ variables. However, because we are integrating either over the $\rho$ density variable or over the particles positions $\v{X}_i$, we can seamlessly do the conversion from the two representations by both switching the integration measure and the probability distribution. Thus, we keep the same notation for the steady-state distribution, the correct one to use is obvious from the context. Now, simply integrating over $\v{x}$ and $\v{y}$ we have 
\begin{multline}
  S_\tau = -\frac{1}{T} \int \prod_i \diff{\v{X}_i} \int \diff{t} \sum_i \Big[ \v{f}(\v{X}_i(t)) \sum_j \nabla V(\v{X}_i-\v{X}_j) + 2 \v{f}(\v{X}_i(t)) \nabla \Upsilon(\v{X}_i(t)) \\
  - T \nabla \v{f}(\v{X}_i(t)) - \v{f}^2(\v{X}_i(t)) \Big] P_{\text{ss}}(\{\v{X}_i\}) \mdot
\end{multline}
In the previous equations $P_\text{ss}$ is the $N$-point density. We introduce the $n$-point density, which we denote $P_n(\v{X}_1, \dots, \v{X}_n)$. It is simply a marginal of the full $N$-point density and can be obtained by integration 
\begin{equation}
  P_n(\v{X}_1, \dots, \v{X}_n) = \frac{1}{(N-n)!} \int \diff{\v{X}_{n+1}} \dots \diff{\v{X}_N} P_N(\v{X}_1, \dots, \v{X}_N) \mdot
\end{equation}
The multiplicity factor is need because we are working with indistinguishable particles. We now marginalized and sum over the particles which only appear in the density and tally the different multiplicity factors, leading to
\begin{equation}
\begin{aligned}
  S_\tau = - & \frac{1}{T} \int \diff{\v{X}_1} \int \diff{t} \Big[ 2 \v{f}(\v{X}_1) \nabla \Upsilon(\v{X}_1) - T \nabla \v{f}(\v{X}_1) - \v{f}^2(\v{X}_1) \Big] P_1(\v{X}_1) \\
  & - \frac{2}{T} \int \diff{\v{X}_1} \diff{\v{X}_2} \int \diff{t} \v{f}(\v{X}_1) \nabla V(\v{X}_1-\v{X}_2) P_2(\v{X}_1,\v{X}_2) \mdot
\end{aligned}
\end{equation} 
We now take the non-conservative force to be a constant drift $\v{w}$. We are interested in the average entropy production rate, which is obtained by simply taking the limit of a short path in time. This gives
\begin{equation}\label{eq:GMP_entropy}
  \dot{S} = \int \diff{\v{X}_1} P_1(\v{X}_1) \Big[ \frac{\v{w}^2}{T} - \frac{2\v{w}}{T} \nabla \Upsilon(\v{X}_1)  \Big] - \int \diff{\v{X}_1} \diff{\v{X}_2} P_2(\v{X}_1,\v{X}_2) \frac{2\v{w}}{T} \nabla V(\v{X}_1-\v{X}_2) \mdot
\end{equation}
This ought to be compared with equation (S-V.109) of \cite{pruessnerFieldTheoriesActive2022}, with the change of notations $\v{X}_i \rightarrow \v{x}_i$, $V \rightarrow U$, $T \rightarrow D$, $P_n \rightarrow \rho_n^{(N)}$. It is immediate to see that all the terms containing a $\v{w}$ factor are an exact match. However, it looks like we are also missing many terms. Interestingly, all the missing terms which are the ones which do not contain the drift $\v{w}$ exactly vanish in the steady state and hence are actually not contributing to the entropy production rate. This point was not discussed in the original paper, so we provide additional details in App. \ref{app:cancelation_fokkerplanck}. This can be actually hinted at by the main expression we derived for the entropy production \eqref{eq:avg_entropy_prod}. Indeed, we can have additional terms in the final result \eqref{eq:GMP_entropy} if we do not use the chain rule to write the first term as the time variation of the free energy. These terms will be exactly the ones vanishing here.

In conclusion, we illustrated that our approach was able to match the one of  \cite{pruessnerFieldTheoriesActive2022}. This was an important test, since Dean's approach, although more direct than the Doi-Peliti, could have had subtleties due to the irregularity of the stochastic equation on the density field. Below, we show another direct test by computing the entropy production at the particles level.

\subsection{Direction computation from Langevin equation at the particles level}\label{subsec:direct_particles}
In the simple case of $N$ particle evolving in a pair-potential, with constant self-propulsion, it is possible to compute the entropy production rate very directly at the level of the particles, without the need to introduce a field theory on the density or to a Doi-Peliti formalism. We write the Langevin equation for the $N$ particles as a vector equation on $\v{x} \in \mathbb{R}^{N}$ with each component the position of one of the particle (we assume to simplify the notation that the system is in 1D, the generalization to higher dimension is straightforward)
\begin{equation}\label{eq:vec_langevin}
\begin{gathered}
    \dot{\v{x}} = \mu \v{F} + \v{\zeta} \\
    \avg{\zeta(\tau) : \zeta(\tau')} = 2 D \delta(\tau - \tau') \mcom
\end{gathered}
\end{equation}
where $\mu$ is the mobility matrix, $D$ the diffusion matrix, $\v{F}$ is the force vector and the angled brackets are average over noise realizations. At equilibrium the diffusion matrix is related to the mobility and the temperature by the famous Einstein relation $D = \mu T$. Because we are interested in out-of-equilibrium cases, there is no reason a priori to relate these quantities.

Using standard methods \cite{cocconiEntropyProductionExactly2020,seifertStochasticThermodynamicsFluctuation2012a}, it is easy to write the OM action associated to the position vector $\v{x} \in \mathbb{R}^N$. The forward path probability is 
\begin{equation}
  \mathcal{P}[\v{x}(\tau) | \v{x}_0] = \mathcal{N} \exp(- \mathcal{A}([\v{x}(\tau)])) \mcom
\end{equation}
the normalization factor $\mathcal{N}$ is constant here, so will not contribute to the entropy production. The action is 
\begin{equation}
  \mathcal{A}([\v{x}(\tau)]) = \int_0^{\tau} \diff{t} \frac{[\dot{\v{x}} - \mu \v{F}]D^{-1}[\dot{\v{x}} - \mu \v{F}]}{4} \mdot
\end{equation}
As usual, when we consider the time-reversed path, we need to switch the convention from Itô to anti-Itô, so the action picks up an extra term: 
\begin{equation}
    \mathcal{A}^R([\v{x}(\tau)]) = \int_0^{\tau} \diff{t} \frac{[\dot{\v{x}} + \mu \v{F}]D^{-1}[\dot{\v{x}} + \mu \v{F}]}{4} + \nabla \cdot [\mu \v{F}] \mdot
\end{equation}
The log of the ratio of path probability reads
\begin{equation}\label{eq:log_ratio_particles}
    \hat{S}_\tau = \log \left( \frac{\mathcal{P}[\v{x}(\tau)]}{\mathcal{P}^R[\v{x}(\tau)]}\right) = \int_0^{\tau} \diff{t} \left( \dot{\v{x}} \cdot D^{-1} \mu \v{F} + \nabla \cdot [\mu \v{F}] \right)  \mdot
\end{equation}
We define the probability current $\v{j} = \mu \v{F} P_N - D \v{\nabla} P_N$, which is defined from the Fokker-Planck equivalent to the Langevin equation~\eqref{eq:vec_langevin}, which is $\partial_{\tau} P_N(\v{x}, \tau) = - \v{\nabla} \cdot \v{j}$. $P_N$ is the $N$-point probability distribution of the particle positions. Now, the entropy production rate can be computed by taking the ensemble average of the log ratio of the path probabilities. Once again, because we work with an Itô description, the average is trivial to compute. 
\begin{equation}
\begin{aligned}
    S_\tau &= \avg{\log \left( \frac{\mathcal{P}[\v{x}(\tau)]}{\mathcal{P}^R[\v{x}(\tau)]}\right)} \\
    &= \int \diff{t} \diff{\v{x}} \left( \mu \v{F} \cdot D^{-1} \mu \v{F} P_N + \nabla \cdot [\mu \v{F}] P_N \right) \\ 
    &= \int \diff{t} \diff{\v{x}} \left( \mu \v{F} \cdot D^{-1} [ \v{j} + D \nabla P_N] + \nabla \cdot [\mu \v{F}] P_N \right) \\ 
    &= \int \diff{t} \diff{\v{x}} \mu \v{F} \cdot D^{-1} \v{j} \mcom
\end{aligned}
\end{equation}
where an integration by parts was used for the last step. By using the explicit form of the force vector and of the probability current one can write the average entropy production as 

\begin{equation}
\begin{aligned}
    \dot{S}[P_N] =\frac{1}{N!} \int \diff{x_1} \dots \diff{x_N} P_N(x_1,\dots,x_N) \sum_i \left( \frac{F_i^2}{T} + \partial_{x_i} F_i \right)\mdot
  \end{aligned}
\end{equation}

From this equation is then straightforward to make the connection with the expression obtained from Dean's equation. The additional steps of the derivation of the equivalence can be found in App. \ref{app:equivalence_langevin}.

This result provides a further test of our framework. In the next section, after this positive test, we apply and extend our approach to other cases, including ones that do not have an exact representation in terms of particles systems as they result from a coarse-grain procedure.

\section{Extensions and applications}

\subsection{Entropy production for non-reciprocal interactions}\label{subsec:non_reciprocal_inter}
The results of the previous sections, can be quite easily generalised and applied to the case where the interactions between the particles are non-reciprocal. This is an important case for active matter because often the multi-particles interactions terms are not direct physical interactions (which usually respect Newton third law and therefore are reciprocal), but rather mediated interactions originated for instance from quorum sensing or chemotaxis \cite{dinelliNonreciprocityScalesActive2023,gnanCriticalBehaviorQuorumsensing2022}. Non-reciprocal interactions are also central in biological process, where the individual constituents are complex entities (cells, animals) with simple interactions modeling very complicated physicochemical processes. Hence, there are a priori no constraints on how the different species interact and different choices leads to drastically different dynamics, such as predator-prey versus prey-prey cohabitation \cite{rednerCaptureLambDiffusing1999,meredithPredatorPreyInteractions2020}.

A typical starting point for these settings is to consider $M$ particles of one specie and $N$ of another and have them interact between species in a non-reciprocal manner \cite{ivlevStatisticalMechanicsWhere2015,alstonIntermittentAttractiveInteractions2022}. If we add a specie dependent self-propulsion, the set of couples Langevin equations is
\begin{equation}
\begin{gathered}
    \dot{X}_i = u_1 - \sum_{j=1}^{N} V_1'(X_i - Y_j) + \xi_i \\
    \dot{Y}_j = u_2 - \sum_{i=1}^{M} V_2'(Y_j - X_i) + \eta_j \\
    \avg{\xi_i(t) \xi_j(t')} = 2 D_1 \delta_{ij} \delta(t-t') \\
    \avg{\eta_i(t) \eta_j(t')} = 2 D_2 \delta_{ij} \delta(t-t') \mdot
\end{gathered}
\end{equation}
The interactions are non-reciprocal because $V_1(x-y) \neq -V_2(y-x)$. We could generalise the setup even more by adding interaction between particles of the same specie without difficulty.

We then define the stochastic densities associated with each species
\begin{equation}
    \rho_1(x,t) = \sum_{i=1}^M \delta(x - X_i(t)), \qquad \qquad \rho_2(y,t) = \sum_{j=1}^N \delta(y - Y_j(t)) \mdot
\end{equation}
From the Langevin equations, we can in the usual manner derive the associated evolution equations for the densities 
\begin{equation}
\begin{gathered}
    \partial_t \rho_1 = \nabla( - \rho_1 u_1 + \rho_1 \nabla V_1 \ast \rho_2 + D_1 \nabla \rho_1) + \nabla( \sqrt{2D_1 \rho_1} \xi) \\
    \partial_t \rho_2 = \nabla( - \rho_2 u_2 + \rho_2 \nabla V_2 \ast \rho_1 + D_2 \nabla \rho_2) + \nabla( \sqrt{2D_2 \rho_2} \eta) \\
    \avg{\xi^\alpha(t) \xi^\beta(t')} = \delta^{\alpha \beta} \delta(t-t'), \qquad \avg{\eta^\alpha(t) \eta^\beta(t')} = \delta^{\alpha \beta} \delta(t-t')
\end{gathered}
\end{equation}
The associated OM action is then simply obtained as the sum of the two actions, leading directly to the log ratio of the path probabilities
\begin{multline}
    \log \left( \frac{P}{P^R} \right) = \int \diff{t} \diff{x} \Bigg[ -\frac{1}{D_1}(\nabla^{-1} \partial_t \rho_1)(u_1 - \nabla V_1 \ast \rho_2 - \frac{D_1}{\rho_1} \nabla \rho_1) \\
    -\frac{1}{D_2}(\nabla^{-1} \partial_t \rho_2)(u_2 - \nabla V_2 \ast \rho_1 - \frac{D_2}{\rho_2} \nabla \rho_2)  \Bigg] \mdot
\end{multline}
As a follow-up of \cite{pruessnerFieldTheoriesActive2022}, in \cite{zhangEntropyProductionNonreciprocal2023} the authors apply the Doi-Peliti approach to non-reciprocal interactions. Similarly to what we did in previous sections, it is possible to recover exactly the same result using the computation we just presented. As done above, we have to replace the different densities by their explicit expressions as sum of Dirac delta functions. Then performing the integrations on the position variables and the sum over the different particles, we recover the expression found in the supplementary material of \cite{zhangEntropyProductionNonreciprocal2023}. This computation is technically similar to what in the previous section and in this setup, every manipulation go through without difficulties. The average entropy production rate is
\begin{multline}
    \dot{S} = \int \diff{x} \diff{y} \bigg[ - \nabla^2 V_1(x - y) + \frac{1}{D_1}\Big( \frac{u_1}{N} - \nabla V_1 (x-y) \Big)^2 - \nabla^2 V_2 (y-x) + \frac{1}{D_2} \Big( \frac{u_2}{M} - \nabla V_2(y-x) \Big)^2 \bigg] P_{1,1}^{M,N}(x,y) \\
    + \int \diff{x} \diff{y} \diff{y'} \frac{1}{D_1} \Big( \frac{u_1}{N} - \nabla V_1(x-y) \Big) \Big( \frac{u_1}{N} - \nabla V_1(x-y') \Big) P_{1,2}^{M,N}(x,y,y') \\
    \int \diff{x} \diff{x'} \diff{y} \frac{1}{D_2} \Big( \frac{u_2}{M} - \nabla V_2(y-x) \Big) \Big( \frac{u_2}{M} - \nabla V_2(y-x') \Big) P_{2,1}^{M,N}(x,x',y) \mdot
\end{multline}
A usual $P_{m,n}^{M,N}$ is the $(m+n)$-point correlation function of the $(M+N)$ particles system.

\section{Generalization to space and density dependent diffusivity}\label{sec:entropy_generalize}
In this section we generalised the result obtained in Sect. \ref{sec:entropy_DDFT} to a more general setup where the diffusivity is no longer a constant. We consider a diffusion coefficient which depends explicitly on space and also on a full functional dependence on the density field. This generalization is for instance necessary if one wants to study an ensemble of many Active Brownian Particles (ABPs) or Run-and-Tumble Particles (RTPs) \cite{tailleurStatisticalMechanicsInteracting2008,catesWhenAreActive2013}. The derivation of the density evolution starting from the dynamical evolution of individual particles is done in details in \cite{solonActiveBrownianParticles2015}. Note that, as Dean's equation, this is an exact equation for the density field and no coarse-grain is involved.

This generalization leads to results resembling the ones obtained for coarse-grained DDFT. However, due to the field dependence of the diffusivity, additional terms needs to be added. Technical details are also more involved and require a careful treatment. The main point of attention is that with this added complexity, it is now necessary to correctly account for what is called the spurious drift, in the evolution equation and in the expression for the OM action. This spurious drift is a general feature of SPDE with multiplicative noise \cite{gardinerHandbookStochasticMethods1994,cugliandoloRulesCalculusPath2017,lauStatedependentDiffusionThermodynamic2007}. It is an additional drift force, which depends on the specific time discretization used. One has to take it into account so that the system correctly reaches equilibrium if the force field is conservative. As we will explain in detail, correctly accounting for this spurious drift, as well as the normalization factors in the OM action, is necessary to obtain the correct final expression for the entropy production. 

We consider the following evolution equation for the density, interpreted in the Itô sense \cite{solonActiveBrownianParticles2015}:
\begin{equation}\label{eq:FHD_ito}
  \dot{\rho} = -\nabla \Big( \v{F}(\v{x},[\rho]) \rho - D(\v{x},[\rho]) \nabla \rho + \rho (\nabla_{\v{x}} \frac{\delta}{\delta \rho(\v{x})})D(\v{x},[\rho])  + (2\rho D(\v{x},[\rho]))^{1/2} \v{\eta} \Big) \mdot
\end{equation}
$\v{F}(\v{x},[\rho])$ is the generic deterministic force applied on the system, $\v{\eta}$ is a white noise field and $D(\v{x}, [\rho])$ is the diffusivity, with its general dependence on the field and position explicitly stated. To simplify notations, we may sometimes omit to specify dependence of some of the terms by replacing $(\v{x},[\rho])$ by $[\rho]$ or dropping it altogether. Compared to \eqref{eq:dean_outeq}, we have an additional term -- the third term on the right-hand side of \eqref{eq:FHD_ito} which is the spurious drift. When the diffusivity is a constant, this term is obviously zero, so this new equation reduces to the simpler case studied above. To recover exactly Eq.~\eqref{eq:dean_outeq}, the deterministic force also needs to take the form $\v{F}[\rho] = \v{f} - D \nabla \frac{\delta \mathcal{F}_\text{ex}}{\delta \rho}$ where $\mathcal{F}_\text{ex}$ is the excess free energy. The total free energy is then $\mathcal{F}[\rho] = \mathcal{F}_\text{ex}[\rho] + \int \diff{\v{x}} \rho \log (\rho-1)$. 

As before, in order to derive the entropy production, the first step is to obtain the forward and backward OM actions. By repeating and generalizing the computations made for Dean's equation, we find that they are respectively given by: 
\begin{equation}\label{eq:actions_FHD}
\begin{aligned}
  \mathcal{A}[\rho] &=  \int \diff{t} \diff{\v{x}} \frac{1}{4 D \rho} \left[ \nabla^{-1} \left( \dot{\rho} + \nabla \left( \v{F} \rho - D \nabla \rho + \rho \left( \nabla \frac{\delta}{\delta \rho} \right) D \right) \right) \right]^2 \\
  \mathcal{A}^R[\rho] &=  \int \diff{t} \diff{\v{x}} \frac{1}{4 D \rho} \left[ \nabla^{-1} \left( -\dot{\rho} + \nabla \left( \v{F} \rho - D \nabla \rho - \rho \left( \nabla \frac{\delta}{\delta \rho} \right) D \right) \right) \right]^2 + \mathcal{A}_J[\rho] \mdot
\end{aligned}
\end{equation}
A crucial point in the expressions above, is that when adding the contributions from forward and backward trajectories, as before, we need to switch the discretization convention from anti-Itô to Itô for the backward contribution. Doing so, the spurious drift needs to be correctly accounted for. It turns out that, as we show in the Appendix, this leads to a change of sign of the spurious drift term in the action. 

Following this route, it is immediate to compute the entropy production by taking the logarithm of the ratio of the path probabilities 
\begin{equation}
\begin{aligned}
  \hat{S}_\tau[\rho] &= \log \left( \frac{P[\rho]}{P^R[\rho]} \right) \\
  &= -\int \diff{t} \diff{\v{x}} \left(\nabla^{-1} \dot{\rho} \right) \left(\frac{\v{F}}{D} - \frac{\nabla \rho}{\rho} \right) - \int \diff{t} \diff{\v{x}} \left( \frac{\v{F}}{D} - \frac{\nabla \rho}{\rho} \right) \rho \left( \nabla \frac{\delta}{\delta \rho} \right) D \mdot
\end{aligned}
\end{equation}

To continue the computation we need to explicitly compute $\mathcal{A}_J$, which again requires to correctly discretize the dynamics in both space and time. All the details are laid out in App. \ref{app:technical_details}. The final expression is
\begin{equation}\label{eq:FHD_entropy}
\begin{split}
  \hat{S}_\tau[\rho] = -\int \diff{t} \diff{x} \left( \nabla^{-1} \dot{\rho} \right) \frac{\v{F}(\v{x},[\rho])}{D(\v{x},[\rho])} - \int \diff{\v{x}} \diff{t} \frac{\v{F}(\v{x},[\rho])}{D(\v{x},[\rho])} \rho(\v{x},t) \left( \nabla_{\v{x}} \frac{\delta}{\delta \rho(\v{x})} \right) D(\v{x},[\rho]) \\
  + \int \diff{t} \diff{\v{x}} \rho(\v{x},t) \left( \nabla_{\v{x}} \frac{\delta}{\delta \rho(\v{x})} \right) \v{F}(\v{x},[\rho]) - \Delta \mathcal{F}_\text{log} \mdot
\end{split}
\end{equation}
Here, $\Delta \mathcal{F}_\text{log} = \mathcal{F}_\text{log}(\tau) - \mathcal{F}_\text{log}(0)$ with $\mathcal{F}_\text{log}(t) = \int \diff{\v{x}} \rho(\v{x},t) \log \rho(\v{x},t)$. Compared to the simpler case studied in Sect. \ref{sec:entropy_DDFT}, we have two additional terms. It is immediate that these two terms vanish when we consider a diffusivity and a non-conservative force which do not depend on the density which means we recover simpler result of the previous section. 

The derivation of \eqref{eq:FHD_entropy} was made using Itô calculus, so it is straight forward to compute the ensemble average of the entropy production. In the steady-state, the $\Delta \mathcal{F}_\text{log}$ term and the ratio of normalization factors do not contribute (see \ref{app:normalization_factor} for details), so the ensemble averaged entropy production rate is  
\begin{equation}\label{eq:FHD_entropy_avg}
  \dot{S} = \big< \dot{\hat{S}} \big> = \int \difff{\rho} \int \diff{\v{x}} P_\text{ss}[\rho] \left[ \frac{\v{F}^2 \rho}{D} - \v{F} \nabla \rho +  \rho \left( \nabla \frac{\delta}{\delta \rho} \right) \v{F} \right] \mcom
\end{equation}
where $P_\text{ss}[\rho]$ is the steady-state distribution of the density. We note that in Eq.~\eqref{eq:FHD_entropy_avg} terms stemming from the spurious drift contribution compensate non trivially. This is reassuring because once averaged, the entropy production rate is a physical quantity which should not depend on the discretization choice made to describe the underlying physical process. Still, it was crucial to includes this spurious drift and handle it carefully in order to obtain the correct final result.

\section{Fluctuating Hydrodynamics}
In this section we briefly describe the application of the results of the previous section to Fluctuating Hydrodynamics (FHD) type equation. This application explains how to apply our framework to cases in which DDFT corresponds to coarse-grained stochastic equations. FHD can be obtained by coarse-graining temporal and spatial scales. Through an appropriate coarse-graining procedure one can obtain a description of the system dynamics in terms of the density field with a noise term representing the projection of fast degrees of freedom. A standard form of FHD \cite{dasFluctuatingNonlinearHydrodynamics1986,catesWhenAreActive2013} is an equation which reads
\begin{equation}
  \dot{\rho} = -\nabla \Big( \v{F}(\v{x},[\rho]) \rho - D(\v{x},[\rho]) \nabla \rho + (2\rho D(\v{x},[\rho]))^{1/2} \v{\eta} \Big) \mdot
\end{equation}
If we compare this with Eq.~\eqref{eq:FHD_ito}, we see that the term originating from the spurious drift is absent. One way to understand this is to recall that the density emerges from a coarse-graining procedure. Then it is natural to consider that the diffusivity has the form of an average of the influence of all the particles contained in a small volume around each point. The simplest case of density dependence diffusion coefficient is a linear expression for $D$ such as 
\begin{equation}
  D(\v{x}, [\rho]) = \int \diff{\v{x'}} K(\v{x}-\v{x'}) \rho(\v{x'}) \mcom
\end{equation}
where $K$ is some kernel implementing local averaging\footnote{With this form, one can show with mild assumption on the kernel, that even starting from an exact equation, the spurious drift term vanishes. In fact, \begin{equation}
\begin{aligned}
  \left( \nabla_{\v{x}} \frac{\delta}{\delta \rho(\v{x})} \right) D &= \int \diff{\v{x'}} \left[ \nabla_{\v{x}} \frac{\delta \rho(\v{x'})}{\delta \rho(\v{x})} \right] K(\v{x}-\v{x'}) \\
  &= \int \diff{\v{x'}} [\nabla_{\v{x}} \delta(\v{x'}-\v{x})] K(\v{x}-\v{x'}) \\
  &= \int \diff{\v{x'}} \delta(\v{x'}-\v{x}) \nabla_{\v{x'}} [K(\v{x}-\v{x'})] \\
  &= - \nabla K(0) \mcom
\end{aligned}
\end{equation}
using $\nabla_{\v{x}} \delta(\v{x'}-\v{x}) = -\nabla_{\v{x'}}(\v{x}-\v{x'})$ and an integration by parts. Thus, this vanishes for a symmetric kernel, which is a natural choice.
}. 
 The stochastic entropy production is then a bit simpler than in the microscopic case described above. It reads:
\begin{equation}
  \hat{S}_\tau[\rho] = -\int \diff{t} \diff{x} \left( \nabla^{-1} \dot{\rho} \right) \frac{\v{F}(\v{x},[\rho])}{D(\v{x},[\rho])} + \int \diff{t} \diff{\v{x}} \rho(\v{x},t) \left( \nabla_{\v{x}} \frac{\delta}{\delta \rho(\v{x})} \right) \v{F}(\v{x},[\rho]) - \Delta \mathcal{F}_\text{log} \mdot
\end{equation}
When averaging, one finds exactly the result for the microscopic theory discussed above (since the spurious drift contributions drops out by averaging). This is interesting because it means that while the contributions stemming from $\left( \nabla_{\v{x}} \frac{\delta}{\delta \rho(\v{x})} D \right)$ produce entropy at the level of individual trajectories, all these contributions add up and cancel in average.

\section{Conclusions}

In this work we derived an expression for the entropy production rate of non-equilibrium systems whose dynamics is described by a density field theory. Our method allowed to describe the stochastic realization-dependent entropy production rate, as well as the steady state entropy production rate obtained by taking an ensemble average over trajectories. We considered several starting points for the derivation, depending on the specific form of the evolution equation for the density. The most general result we obtained in Sect. \ref{sec:entropy_generalize}, for Fluctuating Hydrodynamics like theories, where the diffusivity is allowed to depend both on space explicitly and on the density field. We also described in Sect. \ref{sec:entropy_DDFT} simpler setups, namely some form of Dynamical Density Functional Theory which results from FHD by fixing the diffusivity to a constant, hence significantly simplifying the computations and the final result. 

We compared in Sect. \ref{subsec:Doi-Peliti} the result we obtain in the case of $N$ particles evolving with a constant drift and pair potential and an external field, to the result obtained by the authors in \cite{pruessnerFieldTheoriesActive2022}, where they used the powerful Doi-Peliti field theory approach. As we show in details, both techniques leads to the same final result.

The possibility of deriving an expression for entropy production directly from a density field theory \`a la Dean was not obvious a priori; in fact, one might have anticipated inherent difficulties due to the irregular nature of the density field. Nevertheless, the interest in such an approach was working directly with the density field offers the benefit of more transparent and straightforward derivations. Our results demonstrate the feasibility of this approach, thereby paving the way for its application to broader contexts. A natural extension would be to consider fluctuating hydrodynamics field theories, where both the density and momentum fields are involved such as e.g. \cite{dasFluctuatingNonlinearHydrodynamics1986}.

\section{Acknowledgements}
GB thanks Rosalba Garcia-Millan for interesting discussions at the beginning of this work.

\begin{appendices}
\numberwithin{equation}{section}
  
\section{Derivation of Dean equation with non-conserved force}
In this section we briefly provide the derivation of Dean equation starting from the set of Langevin equation~\eqref{eq:langevin} describing the evolution of $N$ particle, coupled by a pair-potential $V$, evolving in a non-conserved force $\v{f}$.\footnote{Note that, even though we could in principle start from a non-conservative force which is different for each particle, so $\v{f}_i$ for particle $i$, this would not allow us to derive an equation for the total density $\rho$. This is expected because to describe a system of particles with a single density field, it is necessary that the particles are indistinguishable. And for this to be true, the non-conserved force cannot possibly be different for each particle, otherwise there would be a way to tell them apart. This comes up explicitly in the following derivation, when summing over all particles to make the total density appear, it would not work if the non-conservative force had an explicit $i$ dependence.} This derivation follows closely the seminal one done by Dean \cite{deanLangevinEquationDensity1996a}. Define the density associated to particle $i$ by
\begin{equation}
  \rho_i(\v{x},t) = \delta(\v{x}-\v{X}_i(t)) \mcom
\end{equation}
and the total density 
\begin{equation}
  \rho(\v{x},t) = \sum_i^N \rho_i(\v{x},t) \mdot
\end{equation}
We can write an arbitrary function of the field $h$ as 
\begin{equation}
  h(\v{X}_i(t)) = \int \diff{\v{x}} \rho_i(\v{x},t) h(\v{x}) \mcom
\end{equation}
and its time derivative as 
\begin{equation}
  \frac{\diff{h}(\v{X}_i)}{\diff{t}} = \int \diff{\v{x}} \frac{\partial \rho_i}{\partial t}(\v{x},t) h(\v{x}) \mdot
\end{equation}
This time derivative can also be computed using Itô's lemma and the Langevin equation~\eqref{eq:langevin}, leading to
\begin{equation}
\begin{aligned}
  \frac{\diff{h}(\v{X}_i)}{\diff{t}} &= \nabla h(\v{X}_i) \cdot \Big[ \v{\eta}_i + \v{f}(\v{X}_i) - \sum_j \nabla V (\v{X}_i - \v{X}_j)  + T \nabla^2 h(\v{X}_i) \Big] \\
  &= \int \diff{\v{x}} \rho_i(\v{x},t) \Big[ \nabla h(\v{x}) \cdot \v{\eta}_i(t) + \nabla h(\v{x}) \cdot \v{f}(\v{x}) - \nabla h(\v{x}) \cdot \sum_j \nabla V(\v{x}-\v{X}_j) + T \nabla^2 h(\v{x}) \Big] \\
  &= \int \diff{\v{x}} h(\v{x}) \Big[ - \nabla \cdot (\rho_i(\v{x},t) \v{\eta}_i(t)) - \nabla \cdot (\rho_i(\v{x},t) \v{f}(\v{x})) \\
  & \hspace{5.8cm}+ \nabla \cdot (\rho_i(\v{x},t) \sum_j \nabla V(\v{x}-\v{X}_j)) + T \nabla^2 \rho_i(\v{x},t)  \Big] \mdot
\end{aligned}
\end{equation}
The relation above is true for any function $h$. So it implies 
\begin{equation}
  \frac{\partial \rho_i}{\partial t} = - \nabla \cdot (\rho_i(\v{x},t) \v{\eta}_i(t)) - \nabla \cdot (\rho_i(\v{x},t) \v{f}(\v{x})) + \nabla \cdot (\rho_i(\v{x},t) \sum_j \nabla V(\v{x}-\v{X}_j(t))) + T \nabla^2 \rho_i(\v{x},t) \mdot
\end{equation}
Summing over $i$ leads to an equation for the total density
\begin{equation}
  \frac{\partial \rho}{\partial t} = - \sum_i \nabla \cdot (\rho_i(\v{x},t) \v{\eta}_i(t)) - \nabla \cdot (\rho(\v{x},t) \v{f}(\v{x})) + \nabla \cdot (\rho(\v{x},t) \sum_j \nabla V(\v{x}-\v{X}_j(t))) + T \nabla^2 \rho(\v{x},t) \mdot
\end{equation}
To close the equation, we need to define a new noise field with equivalent statistics. We define 
\begin{equation}
  \xi(\v{x},t) = - \sum_i \nabla \cdot (\rho_i(\v{x},t) \v{\eta}_i(t)) \mdot
\end{equation}
It is still a Gaussian noise (as a sum of Gaussian random variable), with correlation
\begin{equation}
\begin{aligned}
  \avg{\xi(\v{x},t) \xi(\v{y},t')} &= 2T \delta(t-t') \sum_i \nabla_{\v{x}} \cdot \nabla_{\v{y}} (\rho_i(\v{x},t) \rho_i(\v{y},t')) \\
  &= 2T \delta(t-t') \nabla_{\v{x}} \cdot \nabla_{\v{y}} (\delta(\v{x}-\v{y}) \rho(\v{x},t)) \mdot
\end{aligned}
\end{equation}
We redefine the noise field as 
\begin{equation}
  \xi'(\v{x},t) = \nabla \cdot (\v{\eta}(\v{x},t) (2T\rho(\v{x},t))^{\nicehalf}) 
\end{equation} 
with the uncorrelated white noise field $\eta$
\begin{equation}
  \avg{\eta^\alpha(\v{x},t) \eta^\beta(\v{y},t')} = \delta(t-t') \delta(\v{x}-\v{y}) \delta^{\alpha \beta} \mdot
\end{equation}
It is easy to show that $\xi$ and $\xi'$ have the same statistics, so for our purpose are equivalent. This leads to the Dean equation with added non-conservative force \eqref{eq:dean_outeq} of the main text. 

\section{Derivation of the OM action for Dean equation}\label{app:OM_action_proof}
In this section, we sketch the derivation of the Onsager-Machlup path integral associated to Dean equation~\eqref{eq:dean_outeq}. This derivation is only a sketch because, while it leads to the correct quadratic part in the action, it does not let us compute the contributions stemming from the Jacobian, which are necessary to obtain the correct expression for the time-reversed dynamics. 

The starting point is the Dean equation in the following form, where we added a non-conservative force $f$. Note that here $\v{f}$ is not a functional of $\rho$ but simply a function depending on the space coordinates $\v{x}$.  
\begin{subequations}
\begin{gather}
    \partial_t \rho(\v{x},t) = \nabla \cdot  \left(\rho(\v{x},t) (\nabla \frac{\delta \mathcal{F}}{\delta \rho(\v{x})} - \v{f}(\v{x})) \right) + \nabla \cdot ( \rho^{\frac{1}{2}}(\v{x},t) \v{\eta}(\v{x},t)) \\
    \left< \eta^{\alpha}(\v{x},t) \eta^{\beta}(\v{x'},t') \right> = 2T \delta^{\alpha \beta} \delta(\v{x}-\v{x'}) \delta(t-t') \mdot
\end{gather}  
\end{subequations}
We start by formally expressing the average of an observable $\mathcal{O}[\rho]$
\begin{equation}
    \avg{\mathcal{O}} = \int \difff{\rho} \mathcal{O}[\rho] \avg{\delta [ \partial_t \rho - R[\rho, \v{\eta}]]}_{\v{\eta}} \mcom  
\end{equation}
where $R[\rho, \v{\eta}]$ denotes the right-hand side of Dean equation. Hence, the average is
\begin{equation}
    \avg{\mathcal{O}} = \int \difff{\rho} \mathcal{O}[\rho] \avg{\delta \left[ \partial_t \rho - \nabla \rho (\nabla \frac{\delta \mathcal{F}}{\delta \rho} - \v{f}) - \nabla(\rho^{1/2} \v{\eta}) \right]}_{\v{\eta}} \mdot
\end{equation}
We then rewrite the term in the delta to put $\v{\eta}$ on his own (we are not writing any Jacobian factor, which is precisely why this computation is not sufficient on its own)
\begin{equation}
    \avg{\mathcal{O}} = \int \difff{\rho} \mathcal{O}[\rho] \avg{\delta \left[ \frac{1}{\rho^{1/2}} \nabla^{-1} \Big(\partial_t \rho - \nabla \rho (\nabla \frac{\delta \mathcal{F}}{\delta \rho} - \v{f}) \Big) -  \v{\eta} \right]}_{\v{\eta}} \mdot
\end{equation}
Then we use the probability distribution of the noise to perform the average 
\begin{equation}
    P_{\v{\eta}}[\v{\eta}] \propto \exp{\left(-\frac{1}{4T} \int \diff{\v{x}}\diff{t} \v{\eta}^2(\v{x},t)\right)} \mdot
\end{equation}
Integrating the noise, using the delta function leads to 
\begin{subequations}
\begin{align}
     \avg{\mathcal{O}} &= \int \difff{\rho} \mathcal{O}[\rho] \exp{\left( -\frac{1}{4T} \int \diff{\v{x}} \diff{t} \left( \frac{1}{\rho^{1/2}} \nabla^{-1} (\partial_t \rho - \nabla \rho (\nabla \frac{\delta \mathcal{F}}{\delta \rho} - \v{f}))  \right)^2 \right)} \\
     &= \int \difff{\rho} \mathcal{O}[\rho] \exp{\left( -\frac{1}{4T} \int \diff{\v{x}} \diff{t} \left( \frac{1}{\rho} \nabla^{-1} \partial_t \rho - \nabla \frac{\delta \mathcal{F}}{\delta \rho} + \v{f}  \right) \left( \nabla^{-1} \partial_t \rho - \rho(\nabla \frac{\delta \mathcal{F}}{\delta \rho} - \v{f}) \right) \right) } \\
     &=  \int \difff{\rho} \mathcal{O}[\rho] e^{-\mathcal{A}[\rho]} \mcom
\end{align}
\end{subequations}
with the OM action 
\begin{equation}
    \mathcal{A}[\rho] = \frac{1}{4T} \int \diff{\v{x}}\diff{t} \left( \frac{1}{\rho} \nabla^{-1} \partial_t \rho - \nabla \frac{\delta \mathcal{F}}{\delta \rho} + \v{f} \right) \left( \nabla^{-1} \partial_t \rho - \rho(\nabla \frac{\delta \mathcal{F}}{\delta \rho} - \v{f}) \right) \mdot
\end{equation}

Note that in this section we did not explicitly write the normalization factor which appears in the main text. It was absorbed in the field integration measure.

\section{Technical details on the computation of the entropy production}\label{app:technical_details}
This section gather most of the technical details required to compute the entropy production in the most general we consider, which is presented in Sect. \ref{sec:entropy_generalize}. First we remind some well-known issues related to interpreting Stochastic Partial Differential Equations (SPDEs) with multiplicative noise, notably the necessity to modify the drift term in a discretization dependent manner in order to reach the correct equilibrium distribution when the deterministic force is conservative. Then we compute the dynamical action for the forward and backward process, discussing in detail the importance of the discretization dependent terms coming from the Jacobian and the spurious drift. Leveraging this, let us write the general form of the entropy production \eqref{eq:FHD_entropy}.

\subsection{Discretization of dynamics and spurious drift}
In this subsection we state some well-known facts about the need of a so-called ``spurious drift'' in the dynamics to correctly recover the equilibrium distribution. Note that the name spurious drift is a misnomer. It is crucial to correctly add this term to the drift if one wants to describe the same stochastic process in different discretization. To make the connection clearer with the usual literature, such as \cite{riskenFokkerPlanckEquationMethods1996,garcia-palaciosIntroductionTheoryStochastic2007,lauStatedependentDiffusionThermodynamic2007}, we use standard notations. 

We consider a vector-valued stochastic process $\v{x} = (x_1,\dots,x_d)$, which coordinates evolve according to 
\begin{equation}
  \dot{x}_i = A_i(\v{x}) + \sum_j B_{ij}(\v{x}) \eta_j 
\end{equation}
with $\eta_j$ a standard white noise $\avg{\eta_i(t) \eta_j(t')} = \delta_{ij} \delta(t-t')$. The Fokker-Planck equation associated to this Langevin equation is in general 
\begin{equation}
  \frac{\partial P}{\partial t} = - \sum_i \frac{\partial}{\partial x_i} \Big[ a_i^{(1)}(\v{x},t) P \Big] + \frac{1}{2} \sum_{i,j} \frac{\partial^2}{\partial x_i \partial x_j} \Big[a_{ij}^{(2)}(\v{x},t) P \Big] \mdot
\end{equation}
The coefficients $a_i^{(1)}(\v{x},t)$ and $a_{ij}^{(2)}(\v{x},t)$ are the jump moments of the process. They explicitly depend on the chosen time discretization of the Langevin equation (which are not equivalent due to the fact that the noise amplitude $B_{ij}(\v{x})$ is field dependent). For a chosen discretization, these jump moments can be computed using 
\begin{equation}
\begin{gathered}
  a_i^{(1)}(\v{x},t) = \lim_{\Dt \to 0} \frac{1}{\Dt} \avg{\Dx_i}   \\
  a_{ij}^{(2)}(\v{x},t) = \lim_{\Dt \to 0} \frac{1}{\Dt} \avg{\Dx_i \Dx_j}
\end{gathered}
\end{equation}
where $\Dx_i$ is the increment of the process between two time steps $\Delta x_i = x_i^{n+1} - x_i^n$, where $n$ is the time discretization index. We consider a general $\alpha$-discretization of the stochastic equation $\v{\bar{x}}^n = \v{x}^n + \alpha \v{\Dx}^n$. We will drop this $n$ in the rest of this section, keeping in mind that when we write $\v{x}$, we mean $\v{x}^n$, the value of the process at the beginning of the time step. Then, using a Taylor expansion to the necessary order we get 
\begin{equation}
\begin{aligned}
  \Dx_i &= \Dt A_i(\bar{x}) + \sqrt{\Dt} \sum_j B_{ij}(\v{\bar{x}}) \eta_j \\
  &= \Dt A_i(\v{x} + \alpha \v{\Dx} ) + \sqrt{\Dt} \sum_j B_{ij}(\v{x}+\alpha \v{\Dx}) \eta_j \\
  &= \Dt A_i(\v{x}) + \sqrt{\Dt} \sum_j B_{ij}(\v{x}) \eta_j + \alpha \sqrt{\Dt} \sum_{j,k} \frac{\partial B_{ij}}{\partial x_k}(\v{x}) \Dx_k \eta_j + O(\Dt^{3/2}) \\
  &= \Dt A_i(\v{x}) + \sqrt{\Dt} \sum_j B_{ij}(\v{x}) \eta_j + \alpha \Dt \sum_{j,k,l} \frac{\partial B_{ij}}{\partial x_k}(\v{x}) \eta_j B_{kl}(\v{x}) \eta_l + O(\Dt^{3/2}) \mdot
\end{aligned}
\end{equation}
Averaging over the noise distribution results in 
\begin{equation}
\begin{gathered}
  \avg{\Dx_i} = \Dt A_i(\v{x}) + \alpha \Dt \sum_{j,k} B_{kj}(\v{x}) \frac{\partial B_{ij}}{\partial x_k}(\v{x}) \\
  \avg{\Dx_i \Dx_j} = \Dt B_{ik}(\v{x}) B_{jk}(\v{x}) \mdot
\end{gathered}
\end{equation}
The $\alpha$-discretized Fokker-Planck equation is finally
\begin{equation}
  \frac{\partial P}{\partial t} = - \sum_{i} \frac{\partial}{\partial x_i} \bigg[ \Big( A_i(\v{x}) + \alpha \sum_{j,k} B_{kj}(\v{x}) \frac{\partial B_{ij}}{\partial x_k}(\v{x}) \Big) P \bigg] + \frac{1}{2} \sum_{i,j,k} \frac{\partial^2}{\partial x_i \partial x_j} \bigg[ B_{ik}(\v{x})B_{jk}(\v{x}) P \bigg] \mdot
\end{equation} 
Setting $\alpha=0$ we recover the well-known Itô form of the Fokker-Planck equation. To derive the expression for the spurious drift, we impose that at long times $P$ reaches a state of thermal equilibrium (we set $T=1$), proportional to $\exp(-\mathcal{H})$. If we set 
\begin{equation}\label{eq:force_spurious_general}
  A_i(\v{x}) = - \frac{1}{2} \sum_{j,k} B_{ik} B_{jk} \frac{\partial \mathcal{H}}{\partial x_j} + \frac{1}{2} \sum_{j,k} \frac{\partial}{\partial x_j} \left( B_{ik} B_{jk} \right) - \alpha \sum_{j,k} B_{kj} \frac{\partial B_{ij}}{\partial x_k}
\end{equation} 
then the Fokker-Planck equation becomes
\begin{multline}
  \frac{\partial P}{\partial t} = \sum_i \frac{\partial}{\partial x_i} \sum_{j,k} \bigg[ \frac{1}{2} B_{ik}B_{jk} \frac{\partial \mathcal{H}}{\partial x_j} -\frac{1}{2} \frac{\partial}{\partial x_j}(B_{ik} B_{jk})P + \alpha B_{kj} \frac{\partial B_{ij}}{\partial x_k} P \\
  - \alpha B_{kj} \frac{\partial B_{ij}}{\partial x_k} P + \frac{1}{2} \frac{\partial}{\partial x_j} (B_{ik} B_{jk})P + \frac{1}{2} B_{ik} B_{jk} \frac{\partial P}{\partial x_j} \bigg] \mdot
\end{multline}
It simplifies to 
\begin{equation}
  \frac{\partial P}{\partial t} = \sum_{i,j,k} \frac{\partial}{\partial x_i} \left[ \frac{1}{2} B_{ik} B_{jk} \left( \frac{\partial \mathcal{H}}{\partial x_j} P + \frac{\partial P}{\partial x_j} \right) \right] \mdot
\end{equation}
Which is indeed solved by the equilibrium distribution $P_\text{eq} \propto \exp(-\mathcal{H})$.

We can now turn back to the equation of FHD \eqref{eq:FHD_ito} in order to show that the third term on the right-hand side is exactly the correct spurious drift. The first step is to discretize the space in the FHD dynamics. To do so, we work in one dimension to make computations clearer, and we define $\rho_i(t) = \rho(x = i \Dx, t)$ where $i = 0,\dots, L-1$, the spatially discretize field and $\Dx$ is the lattice spacing (note the change of notation, $\Dx$ is not anymore a time increment of the stochastic process as we used above). We consider periodic boundary conditions $\rho_L = \rho_0$. The spatial gradient is discretized with a simple mid-point finite difference scheme 
\begin{equation}
  \nabla_i O_i = \frac{1}{2 \Dx}(O_{i+1} - O_{i-1}) \mdot
\end{equation}
The noise amplitude is given by 
\begin{equation}
  B_{ij} = - \frac{1}{\sqrt{2} (\Dx)^{3/2}} ( \delta_{i+1j} - \delta_{i-1j}) \sqrt{D_j \rho_j} \mdot 
\end{equation}
Looking back at \eqref{eq:force_spurious_general}, we compute the last two terms, which represent what we called the spurious drift. The partial derivatives are with respect to the field coordinates $\rho_i$. To lighten notations we write $\partial_{\rho_i} = \frac{\partial}{\partial \rho_i}$. This gives
\begin{equation}
\begin{aligned}
  \frac{1}{2} \sum_{j,k} \frac{\partial}{\partial \rho_j} \left( B_{ik} B_{jk} \right) - \alpha \sum_{j,k} B_{kj} \frac{\partial B_{ij}}{\partial \rho_k} &= - \frac{1}{4(\Dx)^3} \sum_{j,k} \frac{\partial}{\partial \rho_j} \left[(\delta_{i+1k} + \delta_{i-1k})(\delta_{j+1k} - \delta_{j-1k}) D_k \rho_k \right] \\
  &- \frac{\alpha}{2 (\Dx)^3} \sum_{j,k} (\delta_{k+1j} - \delta_{k-1j})(\delta_{i+1j} - \delta_{i-1j}) \sqrt{\rho_j D_j} \frac{\partial}{\partial \rho_k}\left( \sqrt{\rho_j D_j} \right) \\
  &= -\frac{1-\alpha}{4 (\Dx)^3} \left( \rho_{i+1}(\partial_{\rho_i} - \partial_{\rho_{i+2}}) D_{i+1} - \rho_{i-1}(\partial_{\rho_{i-2}} - \partial_{\rho_{i}}) D_{i-1} \right) \\
  &= \frac{1-\alpha}{\Dx} \nabla_i \left( \rho_i \left( \nabla_i \frac{\partial}{\partial \rho_i} \right) D_i \right) \mdot
\end{aligned}
\end{equation}
Where $(\nabla_i \frac{\partial}{\partial \rho_i}) = \frac{1}{2 \Dx} (\partial_{\rho_{i+1}} - \partial_{\rho_{i-1}})$ gives the operator $(\nabla_x \frac{\delta}{\delta \rho(x)})$ in the continuum limit.\footnote{Note that in \cite{solonActiveBrownianParticles2015}, the authors define this operator with the opposite sign of what we state here. As far as we can tell this is a mistake and leads to several inconsistencies in their computation. We believe that the definition given in the present paper is the correct one.} Note that in the expression for the spurious drift, there is a factor $\Dx^{-1}$, which prevents the definition of a proper continuous limit. This is due to the know fact that some quantities one need to manipulate when looking at path integral formulations of field theories do not converge in the continuum. This is why a spatial discretization is crucial. However, in the final physical results, these diverging quantities always compensate to give rise to properly define continuous expressions.  

From this computation, we have the Itô FHD equation 
\begin{equation}
  \dot{\rho} \stackrel{\alpha=0}{=} -\nabla \Big( F(x,[\rho]) \rho - D(x,[\rho]) \nabla \rho + \rho (\nabla_x \frac{\delta}{\delta \rho(x)})D(x,[\rho])  - (2\rho D(x,[\rho]))^{1/2} \eta \Big) \mdot
\end{equation}
and the anti-Itô version
\begin{equation}\label{eq:anti_ito_FHD}
  \dot{\rho} \stackrel{\alpha=1}{=} -\nabla \Big( F(x,[\rho]) \rho - D(x,[\rho]) \nabla \rho + (2\rho D(x,[\rho]))^{1/2} \eta \Big) \mdot
\end{equation}

\subsection{Dynamical action for general time discretization}
In this subsection we compute the dynamical action for a general time discretization. As stated in Sect. \ref{sec:entropy_generalize}, when computing the action for the forward and time-reversed dynamics, it is essential to be careful about the spurious drift and the Jacobian contributions. Correctly taking them into account is essential to derive the correct expression of the entropy production. Again, to be precise it is best to work with the discretized dynamics. 

For the time discretization, we index time by $s=0,\dots,S$, with time step $\Dt$. Different temporal schemes can be used to evaluate the different functions. They are defined by $x_i^{s+\alpha} = \alpha x_i^s + (1-\alpha) x_i^{s+1}$. Usual choices are $\alpha=0$ for Itô, $\alpha=\frac{1}{2}$ for Stratonovich, $\alpha=1$ for anti-Itô. The path probability was fully derived in \cite{lauStatedependentDiffusionThermodynamic2007}, albeit using more of a response field formalism. Performing a simple Gaussian integration, the OM form was written by \cite{catesStochasticHydrodynamicsComplex2022}, which is our starting point to compute the entropy  The path probability is
\begin{equation}
  \mathcal{P}(\{\rho_i^S\}) = \mathcal{N}(\{\rho_i^S\}) e^{- \mathcal{A}(\{ \rho_i^S\})} \mdot
\end{equation}
With the normalization factor equals to 
\begin{equation}
  \mathcal{N}(\{\rho_i^S\}) = \prod_{s=0}^{S-1} \frac{1}{\det (B_{ij}^{s+\alpha})\sqrt{2 \pi \Dt}} \mcom
\end{equation}
and the dynamical action  
\begin{multline}\label{eq:dyn_action}
  \mathcal{A}(\{\rho_i^s\}) = \sum_{s=0}^{S-1} \Dt \bigg\{ \frac{1}{2} \sum_{i,j,k,l,m,n,p} \Big( \frac{\rho_i^{s+1} - \rho_i^{s}}{\Dt} - A_i + \alpha B_{ik} \frac{\partial B_{lk}}{\partial \rho_l}  \Big) (B_{ip} B_{jp})^{-1} \Big( \frac{\rho_j^{s+1} - \rho_j^{s}}{\Dt} - A_j + \alpha B_{jm} \frac{\partial B_{nm}}{\partial \rho_n}  \Big) \\ 
  + \alpha \sum_i \frac{\partial A_i}{\partial \rho_i} + \frac{\alpha^2}{2} \sum_{i,j,k} \Big( \frac{\partial B_{ik}}{\partial \rho_j} \frac{\partial B_{jk}}{\partial \rho_i} - \frac{\partial B_{ik}}{\partial \rho_i} \frac{\partial B_{jk}}{\partial \rho_j}  \Big)   \bigg\} \mdot
\end{multline}
In the previous equation the functions of the field $A_i$ and $B_{ij}$ are evaluated at $\rho^{s+\alpha}$. 

To evaluate the entropy production rate, it is necessary to compute the action of the time-reversed path. We define the time-reversed field by ${\rho_i^s}^{R} = \rho_i^{S-s}$. One crucial point is that when considering the reversed dynamics, the time-discretization convention needs to be changed from $\alpha$ to $1-\alpha$. So if we start in Itô, we need to evaluate the reversed action in anti-Itô. 

Note that since the field theory conserves the total density, we are working on a functional space in which the field component corresponding to the average value, i.e. the zero frequency Fourier component, is not part of the space. This justifies why the operator $B_{ij}$ does not have a zero mode, and hence a zero determinant. 

\paragraph{Spurious drift}
We now show that the forms of dynamical action and the time-reversed one are indeed given by \eqref{eq:actions_FHD}. First we compute the quadratic term appearing in the action, using the expression we derived for the spurious drift. To stay more general we do not assume that the deterministic force derive form a potential, thus we keep an unspecified form, which we write $C_i$
\begin{equation}
\begin{aligned}
  -A_i + \sum_{k,l} \alpha B_{ik} \frac{\partial B_{lk}}{\partial \rho_l} &= C_i - \frac{1}{2} \sum_{j,k}  \frac{\partial}{\partial \rho_j}(B_{ik} B_{jk}) + \alpha \sum_{j,k} B_{kj} \frac{\partial B_{ij}}{\partial \rho_k} + \alpha \sum_{k,l} B_{ik} \frac{\partial B_{lk}}{\partial \rho_l} \\
  &= C_i + \left( \alpha - \frac{1}{2} \right) \sum_{j,k} \frac{\partial}{\partial \rho_j} (B_{ik} B_{jk}) \\
  &= C_i + \frac{(1- 2\alpha)}{\Dx} \nabla_i \left( \rho_i \left( \nabla_i \frac{\partial}{\partial \rho_i} \right) D_i \right) \mdot
\end{aligned}
\end{equation}
Hence in the continuous limit we clearly see that we indeed have the announced quadratic part in \eqref{eq:actions_FHD} due to the change from $\alpha=0$ in the forward to $\alpha=1$ in the time-reversed. Crucially we see that the sign in front of the spurious drift changes form one case to the other. 
\paragraph{Normalization Factor}\label{app:normalization_factor}
In the computation of the entropy production, the relevant quantity concerning the normalization factor is the logarithm of their ratio. If the forward dynamics is in the Itô discretization, the ratio of the normalization factor can be simplified (expect for the first and last points, the other time points are the same, so most of the factors simplify pairwise, this is however not the case if we consider an arbitrary time discretization) and leads to
\begin{equation}
  \log \left( \frac{\mathcal{N}(\{\rho_i^S\})}{\mathcal{N}(\{{\rho_i^S}^R\})} \right) = \log \left( \frac{\det \left({B}_{ij}^S \right)}{ \det \left({B}_{ij}^0 \right)} \right) \mdot
\end{equation}
Evaluating this term is complicated, but we can use a trick to show that it is zero. In fact, this term is precisely the one that should be canceled by the spurious drift contribution when the latter is present. In the case of Dean's equation the spurious drift is zero as we discussed in the main text, and in consequence this terms has to be zero too. 

In any case, even without resorting to this argument, one can conclude that it does not play any role in the average entropy production since 
\begin{equation}
  \avg{\log \left( \frac{\det \left( {B}_{ij}^S \right)}{ \det \left( {B}_{ij}^0 \right)} \right)}_{ss} = \avg{\det \left( {B}_{ij}^S \right)}_{ss} - \avg{\det \left( {B}_{ij}^0 \right)}_{ss} = 0
\end{equation}
thanks to stationarity.
\paragraph{Rest of Jacobian contribution}
Expect for Itô where $\alpha=0$, it is clear that \eqref{eq:dyn_action} contains additional terms apart from the quadratic part. These terms are denoted $\mathcal{A}_J[\rho]$ in the main text. We compute them explicitly taking the force part we wrote for FHD \eqref{eq:anti_ito_FHD} in the anti-Itô discretization (hence the spurious drift contribution vanishes). We have 
\begin{equation}
  A_i = -\frac{1}{2\Dx} \left(F_{i+1} \rho_{i+1} - F_{i-1} \rho_{i-1} \right) + \frac{1}{(2\Dx)^2} \left(D_{i+1} \rho_{i+2} + D_{i-1}\rho_{i-2} - D_{i+1} \rho_i - D_{i-1} \rho_i \right) \mdot
\end{equation}
Hence, the contribution to the action is 
\begin{align*}\label{eq:jacobian_FHD}
  \sum_i \frac{\partial A_i}{\partial \rho_i} &= -\frac{1}{2\Dx} \sum_i ( \rho_{i+1} \partial_{\rho_i} F_{i+1} + F_{i+1} \delta_{i i+1} - \rho_{i-1} \partial_{\rho_i}F_{i-1} - F_{i-1} \delta_{i i-1} ) \\
  & \quad + \frac{1}{(2\Dx)^2} \sum_i ( \rho_{i+2} \partial_{\rho_i} D_{i+1} + D_{i+1} \delta_{i i+2} + \rho_{i-2} \partial_{\rho_i} D_{i-1} + D_{i-1} \delta_{ii-2} \\
  & \qquad - \rho_i \partial_{\rho_i} D_{i+1} - D_{i+1} - \rho_i \partial_{\rho_i} D_{i-1} - D_{i-1} ) \\
  &= -\frac{1}{2\Dx} \sum_i \rho_i (\partial_{\rho_{i-1}} - \partial_{\rho_{i+1}}) F_i \\
  & \qquad + \frac{1}{(2\Dx)^2} \sum_i ( \rho_{i+1} \partial_{\rho_{i-1}} D_i + \rho_{i-1} \partial_{\rho_{i+1}} D_i - \rho_{i-1} \partial_{\rho_{i-1}} D_i - \rho_{i+1} \partial_{\rho_{i+1}} D_i - 2D_i) \\
  &= -\frac{1}{2\Dx} \sum_i \rho_i (\partial_{\rho_{i-1}} - \partial_{\rho_{i+1}}) F_i + \frac{1}{(2\Dx)^2} \sum_i (\rho_{i+1} - \rho_{i-1})(\partial_{\rho_{i-1}} - \partial_{\rho_{i+1}}) D_i \\
  & \qquad - \frac{2}{(2\Dx)^2} \sum_i D_i \\
  &= \frac{1}{2\Dx} \sum_i \rho_i (\partial_{\rho_{i+1}} - \partial_{\rho_{i-1}}) F_i - \frac{1}{(2\Dx)^2} \sum_i (\rho_{i+1} - \rho_{i-1})(\partial_{\rho_{i+1}} - \partial_{\rho_{i-1}}) D_i \\
  & \qquad - \frac{1}{(2\Dx)^2} \sum_i D_i (\partial_{\rho_{i+1}} - \partial_{\rho_{i-1}})(\rho_{i+1} - \rho_{i-1}) \\
  & \xrightarrow[\Dx \to 0]{} \int \diff{x} \rho \left(\nabla \frac{\delta}{\delta \rho} \right) F - \int \diff{x} (\nabla \rho)\left(\nabla \frac{\delta}{\delta \rho}\right) D - \int \diff{x} D \left( \nabla \frac{\delta}{\delta \rho} \right) \nabla \rho \mcom \stepcounter{equation} \tag{\theequation}
\end{align*}  

The final contribution to the Jacobian is the cross derivative term, composed of two terms
\begin{equation}
\begin{aligned}
  \frac{1}{2} \sum_{i,j,k} \frac{\partial B_{ik}}{\partial \rho_j} \frac{\partial B_{jk}}{\partial \rho_i} &= \frac{1}{2(2\Dx)^3} \sum_{i,j,k} (\delta_{i+1 k} - \delta_{i-1 k})(\delta_{j+1 k} - \delta_{j-1 k}) \\
  & \qquad \times \left[ \sqrt{\frac{\rho_k}{D_k}} \partial_{\rho_j} D_k + \sqrt{\frac{D_k}{\rho_k}} \delta_{kj} \right] \left[  \sqrt{\frac{\rho_k}{D_k}} \partial_{\rho_i} D_k + \sqrt{\frac{D_k}{\rho_k}} \delta_{ki} \right] \\
  &= \frac{1}{2(2\Dx)^3} \sum_i \frac{\rho_i}{D_i} \left[ \frac{\partial D_i}{\partial \rho_{i+1}} - \frac{\partial D_i}{\partial \rho_{i-1}} \right]^2 \mcom
\end{aligned}
\end{equation} 
and 
\begin{equation}
\begin{aligned}
  - \frac{1}{2} \sum_{i,j,k} \frac{\partial B_{ik}}{\partial \rho_i} \frac{\partial B_{jk}}{\partial \rho_j} &= - \frac{1}{2(2\Dx)^3} \sum_{i,j,k} (\delta_{i+1 k} - \delta_{i-1 k})(\delta_{j+1 k} - \delta_{j-1 k}) \\
  & \qquad \times \left[ \sqrt{\frac{\rho_k}{D_k}} \partial_{\rho_i} D_k + \sqrt{\frac{D_k}{\rho_k}} \delta_{ki} \right] \left[  \sqrt{\frac{\rho_k}{D_k}} \partial_{\rho_j} D_k + \sqrt{\frac{D_k}{\rho_k}} \delta_{kj} \right] \\
  &= -\frac{1}{2(2\Dx)^3} \sum_i \frac{\rho_i}{D_i} \left[ \frac{\partial D_i}{\partial \rho_{i+1}} - \frac{\partial D_i}{\partial \rho_{i-1}} \right]^2 \mdot
\end{aligned}
\end{equation}
Hence, it is immediate that
\begin{equation}
  \frac{1}{2} \sum_{i,j,k} \left( \frac{\partial B_{ik}}{\partial \rho_j} \frac{\partial B_{jk}}{\partial \rho_i} - \frac{\partial B_{ik}}{\partial \rho_i} \frac{\partial B_{jk}}{\partial \rho_j} \right) = 0 \mcom
\end{equation}
thus, this cross derivatives term does not contribute.

\subsection{Itô's lemma}
We carried out the different computations starting from an Itô stochastic differential equation. Hence, it is crucial to use Itô's lemma when one needs to calculate the time derivative of a function of the stochastic field. In this subsection we compute the general contributions stemming from Itô's formula 

As a reminder, see for instance \cite{gardinerHandbookStochasticMethods1994}, we consider a set of stochastic variables $\{x_i\}_i$, which in vector notation evolve according to 
\begin{equation}
  \diff{\v{x}} = A(\v{x},t) \diff{t} + B(\v{x},t) \diff{W}(t) \mcom
\end{equation} 
where $\v{x} \in \mathbb{R}^d$, $A: \mathbb{R}^d \times \mathbb{R} \rightarrow \mathbb{R}^d$, $B: \mathbb{R}^d \times \mathbb{R} \rightarrow \mathbb{R}^{d \times d}$ and $\diff{W}$ is a Wiener process.  Let $\phi: \mathbb{R}^d \rightarrow \mathbb{R}$ be a function of the stochastic variables. Then Itô's lemma gives the differential of $\phi$ 
\begin{equation}
  \diff{\phi(\v{x})} = \Big( A_i(\v{x},t) \partial_i \phi(\v{x}) + \frac{1}{2} [B(\v{x},t) B^T(\v{x},t)]_{ij} \partial_i \partial_j \phi(\v{x}) \Big) \diff{t} + B_{ij} \partial_i \phi(\v{x}) \diff{W_j(t)} \mdot
\end{equation}

We now adapt the notation to our setup. The extra term that we need to add to the usual chain rule is, for an arbitrary function of the field $\phi[\rho]$  
\begin{equation}
\begin{aligned}
  \frac{1}{2} \sum_{i,j,k} B_{ik} B_{jk} \partial_{\rho_i} \partial_{\rho_j} \phi &= \frac{1}{4 (\Dx)^3} \sum_{i,j,k} \left( \delta_{i+1k} \delta_{j+1k} + \delta_{i-1k} \delta_{j-1k} \right. \\
  & \qquad \left. - \delta_{i+1k} \delta_{j-1k} - \delta_{i-1k} \delta_{j+1k} \right) D_k \rho_k \partial_{\rho_i} \partial_{\rho_j} \phi \\
  &= \frac{1}{4 (\Dx)^3} \sum_{i} \left( D_{i+1} \rho_{i+1} \partial_{\rho_i} \partial_{\rho_i} \phi + D_{i-1} \rho_{i-1} \partial_{\rho_i} \partial_{\rho_i} \phi \right. \\
  & \qquad \left.- D_{i+1} \rho_{i+1} \partial_{\rho_i} \partial_{\rho_{i+2}} \phi - D_{i-1} \rho_{i-1} \partial_{\rho_i} \partial_{\rho_{i-2}} \phi \right) \\
  &= \frac{1}{4(\Dx)^3} \sum_{i} D_i \rho_i \left( \partial_{\rho_{i-1}} \partial_{\rho_{i-1}} + \partial_{\rho_{i+1}} \partial_{\rho_{i+1}} - 2 \partial_{\rho_{i+1}} \partial_{\rho_{i-1}} \right) \phi \\
  &= \frac{1}{4(\Dx)^3} \sum_{i} D_i \rho_i \left( \partial_{\rho_{i+1}} - \partial_{\rho_{i-1}} \right)^2 \phi \\
\end{aligned}
\end{equation}
In practice we are interested in cases where $\phi[\rho]$ is expressed as the integral of some function of the field
\begin{equation}
  \phi[\rho] = \int \diff{x} \psi(\rho(x))
\end{equation}
with $\psi$ some function of $\rho(x)$. Then discretizing space and applying Itô's lemma gives a term equal to 
\begin{equation}
\begin{aligned}
  \frac{1}{2} \sum_{i,j,k} B_{ik} B_{jk} \partial_{\rho_i} \partial_{\rho_j} \phi &= \frac{1}{4 (\Dx)^3} \sum_i D_i \rho_i (\partial_{\rho_{i+1}} - \partial_{\rho_{i-1}})^2 \Dx \sum_j \psi(\rho_j) \\
  & \xrightarrow[\Dx \to 0]{} \int \diff{x} D \rho \left(\nabla \frac{\delta}{\delta \rho} \right)^2 \psi(\rho(x))
\end{aligned}
\end{equation} 

We can apply this result to $\mathcal{F}_\text{log}[\rho] = \int \diff{y} \rho(y) \log \rho(y)$, it gives
\begin{equation}
  \begin{aligned}
    \int \diff{x} D \rho \left( \nabla \frac{\delta}{\delta \rho} \right)^2 \left(\rho(x) \log \rho(x) \right) &=  \int \diff{x} D \rho \left( \nabla \frac{\delta}{\delta \rho} \right) \nabla \log \rho \\
    &= \int \diff{x} D \left( \nabla \frac{\delta}{\delta \rho} \right) \nabla \rho \mcom
\end{aligned}
\end{equation}
where we used the fact that $\rho$ commutes with the operator $\left( \nabla \frac{\delta}{\delta \rho} \right)$.\footnote{This is obvious from the discrete definition of the operator. Indeed, it is $(2\Dx)^{-1}(\partial_{\rho_{i+1}} - \partial_{\rho_{i-1}})$ so if we apply it to $\rho_i$ it directly vanishes.} We see that this term is exactly found in the drift contribution to the Jacobian, the last term on the last line of \eqref{eq:jacobian_FHD}. This was used in the derivation of the entropy production to write the logarithmic part as a total variation $\Delta \mathcal{F}_\text{log}$.

Bringing all the computations explained in this section together directly leads to the general expression for the entropy production in the continuum \eqref{eq:FHD_entropy}.

\subsection{Vanishing entropy production at equilibrium}
As a sanity check, we quickly demonstrate how the expression for the entropy production we obtained in the general case, gives the correct result when the system is in equilibrium. As already stated, the condition on the force field $\v{F}$ to be at equilibrium is to have 
\begin{equation}
  \v{F}_\text{eq}(\v{x}, [\rho]) = - D(\v{x}, [\rho]) \nabla \frac{\delta \mathcal{F}_\text{ex}}{\delta \rho(\v{x})} \mdot
\end{equation}
where $\mathcal{F}_\text{ex}$ is the excess free-energy. Plugging this expression in \eqref{eq:FHD_entropy} gives
\begin{equation}
\begin{aligned}
  \hat{S}_\tau[\rho] &= \int \diff{t} \diff{x} \left( \nabla^{-1} \dot{\rho} \right) \nabla \frac{\delta \mathcal{F_\text{ex}}}{\delta \rho(\v{x})} + \int \diff{\v{x}} \diff{t} \rho(\v{x},t) \left(\nabla \frac{\delta \mathcal{F}_\text{ex}}{\delta \rho(\v{x})} \right) \left( \nabla_{\v{x}} \frac{\delta}{\delta \rho(\v{x})} \right) D(\v{x},[\rho]) \\
  & \qquad - \int \diff{t} \diff{\v{x}} \rho(\v{x},t) D(\v{x},[\rho]) \left( \nabla_{\v{x}} \frac{\delta}{\delta \rho(\v{x})} \right)^2 \mathcal{F}_\text{ex} - \int \diff{t} \diff{\v{x}} \rho(\v{x},t) \left( \nabla_{\v{x}} \frac{\delta \mathcal{F}_\text{ex}}{\delta \rho(\v{x})} \right) \left( \nabla_{\v{x}} \frac{\delta}{\delta \rho(\v{x})} \right) D(\v{x},[\rho]) \\
  &\qquad - \Delta \mathcal{F}_\text{log} \\
  &= - \Delta \mathcal{F}_\text{ex} - \Delta \mathcal{F}_\text{log} \\
  &= - \Delta \mathcal{F} \mdot
\end{aligned}
\end{equation}
Once again to use Itô's lemma to write the pair of terms as the total time differential of the free energy. In the end, we recover the expected result for the equilibrium situation.

\subsection{Continuous description}
It would be desirable to have a way to perform all the computation in the continuum, leveraging for instance the methods used in \cite{hochbergEffectiveActionStochastic1999,zinn-justinQuantumFieldTheory2021}. The difficult step is to make sense of the functional Jacobian determinant necessary to write down the action. The dynamical equation has several difficult features: it is non-linear, the noise is multiplicative, and the noise amplitude is non-linear in the field. The non-linearity in the deterministic part of the dynamics can be dealt with using a standard approach. However, the non-linearity in the noise amplitude is much more challenging. One way forward could be to use the usual Faddeev-Popov ghosts method to trade the functional determinant for an integration over a Grassmann field and its conjugate. We did not explore this strategy much further, but it is definitely something that would be worth to pursue in a follow-up work.

\subsection{Simplifications for DDFT}\label{app:DDFT_entropy_simplification}
The expression of the entropy production we presented in Sect. \ref{sec:entropy_DDFT} can be directly obtained form the general expression obtained in \ref{sec:entropy_generalize}. In order to do so, the diffusivity $D(x,[\rho])$ is taken as a simple density and space independent constant $T$. The non-conservative force is also turned into a free energy pair potential part and a density independent but space dependent force $f(x)$. This directly leads to the appropriate simplifications to recover the simpler form of the entropy production \eqref{eq:DDFT_entropy_production_stoch}.

\section{Cancelation in the entropy production rate expression}\label{app:cancelation_fokkerplanck}

In the computation by Pruessner and Garcia-Millan (PGM), one of their main result is equation (S-V.109), which gives the entropy production rate of $N$ pair-interacting indistinguishable particles in an external potential with a drift. As a very important check of the approach developed in the present paper, we need to show that we can derive the same final result. One key interrogation raised in our derivation, is that it looked like, while we correctly recovered the term involving the non-equilibrium drift contribution, the other terms were vanishing in the steady state, because we could write them as the time variation of a free energy. When studying more precisely (S-V.109) of PGM, it is not obvious why the corresponding terms are actually not contributing, a thing that was in fact, as far as we can tell, not noticed by the authors in their paper. In this section we show that the terms in fact correctly do not contribute to the entropy production. The key is to use the Fokker-Planck (FP) equation and to manipulate it to recover exactly the terms, which then give a vanishing contribution in the steady state.

We denote the N particles coordinates by $x_i$, $V(x_i - x_j)$ is the pair potential, $\Upsilon(x_i)$ is the external field, $w$ is the drift and $T$ the noise amplitude. They are respectively equal to $x_i$, $U$, $\Upsilon$, $w$ and $D$ in equation (S-V.109). We start from the Fokker-Planck equation for the N-point density $\rho_N^{(N)}(x_1,\dots,x_N)$ for a system of $N$ particles (we drop the $(N)$ to ease notation, only keeping the index denoting the N-point density).
\begin{equation}
  \frac{\partial \rho_N}{\partial t}  = \sum_i \frac{\partial}{\partial x_i} \bigg( \Big[ \sum_j V'(x_i - x_j) + \Upsilon' (x_i) \Big] \rho_N \bigg) + T \sum_i \frac{\partial^2 \rho_N}{\partial x_i^2} \mdot
\end{equation}
The trick is now to multiply the FP equation by 
\begin{equation}
  -\frac{1}{2 T (N-2)!} \Big( V(x_1-x_2) + \frac{2}{N-1} \Upsilon(x_1) \Big)
\end{equation}
and integrate it over the $x_1, \dots, x_N$. Because we consider a steady state, the left-hand side trivially gives zero. The right-hand side will be exactly the term we are after, after proper integration by parts, summations, marginalization over additional position variables and reorganization of dummy variables. We use explicitly the fact that the pair-potential $V$ is symmetric. We have 

\begin{align*}
  &\int \diff{x_{1,\dots,N}} \bigg[ V(x_1 - X_2) + \frac{2}{N-1} \Upsilon(x_1) \bigg] \sum_i \bigg[ \frac{\partial}{\partial x_i} \Big( \sum_j V'(x_i-x_j) \rho_N \Big) + \frac{\partial}{\partial x_i} \Big( \Upsilon'(x_i) \rho_N \Big) \\
  & \hspace{1.4cm} + T \frac{\partial^2}{\partial x_i^2} \rho_N \bigg] \\
  &= \int \diff{x_{1,\dots,N}} \bigg[ -V'(x_1-x_2)V'(x_1-x_2) - V'(x_1-x_2) \sum_{j \neq 2} V'(x_1-x_j) + V'(x_1-x_2)V'(x_2-x_1) \\ 
  & \hspace{1cm} + V'(x_1-x_2) \sum_{j \neq 1} V'(x_2-x_j) - \frac{2}{N-1} \Upsilon'(x_1) \sum_{j \neq 1} V'(x_1-x_j) - V'(x_1-x_2) \Upsilon(x_1) \\
  & \hspace{1cm} + V'(x_1-x_2) \Upsilon'(x_2) - \frac{2}{N-1} \Upsilon'(x_1)^2 + 2T V''(x_1-x_2) + \frac{2}{N-1} T \Upsilon''(x_1) \bigg] \rho_N \stepcounter{equation} \tag{\theequation}  \\
  &= -2(N-2)! \int \diff{x_{1,2}} \Big[V'(x_1-x_2)]^2 \rho_2(x_1,x_2) \\
  & \hspace{1cm} - 2(N-3)!(N-2) \int \diff{x_{1,2,3}} V'(x_1-x_2)V'(x_1-x_3) \rho_3(x_1,x_2,x_3) \\
  & \hspace{1cm} - 4 (N-2)! \int \diff{x_{1,2}} \Upsilon'(x_1) V'(x_1-x_2) \rho_2(x_1,x_2) - 2(N-2)! \int \diff{x_1} \Big[\Upsilon'(x_1)\Big]^2 \rho_1(x_1) \\
  & \hspace{1cm}+ 2T(N-2)! \int \diff{x_{1,2}} V''(x_1-x_2) \rho_2(x_1,x_2) + 2T(N-2)! \int \diff{x_1} \Upsilon''(x_1) \rho_1(x_1) \\
  &= 0 \mdot
\end{align*}
Multiplying this equation by $-\frac{1}{2T(N-2)!}$ shows that terms not proportional to the drift indeed cancel in (S-V.109).

\section{Equivalence of the Langevin computation with other approaches}\label{app:equivalence_langevin}
It is possible to show that the result of Sect. \ref{subsec:direct_particles} exactly matches the one obtained in the same setup by the authors of \cite{pruessnerFieldTheoriesActive2022}, which we discussed at length in Sect. \ref{subsec:Doi-Peliti}. To do so, we define the force on particle $i$ to be $F_i = - \sum_{j \neq i} \partial_i V(x_i - x_j) + w = -\sum_{j \neq i} V'(x_i - x_j) + w $, with $V$ the pair potential and $w$ the self-propulsion. We take $D$ and $\mu$ to be diagonal and related by the Einstein relation $D = \mu T$ to recover the same result. The entropy production rate is then 
\begin{equation}
\begin{aligned}
    \dot{S}[P_N] &= \frac{1}{N!} \int \diff{x_1} \dots \diff{x_N} \frac{\v{F} \cdot \v{j}}{T} \\
    &= \frac{1}{N!} \int \diff{x_1} \dots \diff{x_N} P_N(x_1,\dots,x_N) \sum_i \left( \frac{F_i^2}{T} + \partial_{x_i} F_i \right) \\
    &= \frac{1}{N!} \int \diff{x_1} \dots \diff{x_N} P_N \sum_i \bigg( \frac{w^2}{T} + \frac{1}{T}\sum_{j \neq i}\sum_{k \neq i} V'(x_i-x_j) V'(x_i-x_k) \\
    & \hspace{4.5cm} - \frac{2w}{T} \sum_{j \neq i} V'(x_i-x_j) - \sum_{j \neq i} V''(x_i -x_j) \bigg) \mdot
  \end{aligned}
\end{equation}
The $N!$ factor comes from the probability distribution and is necessary in the case of indistinguishable particle to avoid overcounting. 
We now use the marginalization property of the probability density 
\begin{equation}
    \int \diff{x_n} P_n(x_1, \dots, x_n) = (N - (n-1)) P_{n-1}(x_1,\dots,x_{n-1})
  \end{equation}
to marginalized over the superfluous particles coordinates and, we also split one of the term into two particles and the three particles interactions terms
\begin{equation}
\begin{aligned}
    \dot{S}[\rho_N] &= \sum_i \frac{(N-1)!}{N!} \int \diff{x_i} P_1(x_i) \frac{w^2}{T} \\
    & \hspace{.5cm} + \sum_i \sum_{j \neq i} \frac{(N-2)!}{N!} \int \diff{x_i} \diff{x_j} P_2(x_i,x_j) \bigg[ \frac{1}{T} (V'(x_i-x_j))^2 - \frac{2w}{T} V'(x_i-x_j) - V''(x_i-x_j) \bigg] \\
    & \hspace{.5cm} + \sum_i \sum_{j \neq i } \sum_{k \notin \{i,j\}} \frac{(N-3)!}{N!} \int \diff{x_i} \diff{x_j} \diff{x_k} P_3 (x_i,x_j,x_k) \frac{1}{T} V'(x_i-x_j)V'(x_i-x_k) \\
    &= N\frac{w^2}{T} + \int \diff{x_1} \diff{x_2} P_2(x_1,x_2) \bigg[ \frac{1}{T} (V'(x_1-x_2))^2 - \frac{2w}{T} V'(x_1-x_2) - V''(x_1-x_2) \bigg] \\
    & \hspace{.5cm} + \int \diff{x_1} \diff{x_2} \diff{x_3} P_3(x_1,x_2,x_3) \frac{1}{T} V'(x_1-x_2) V'(x_1-x_3) \mcom
  \end{aligned}
\end{equation}
where we used
\begin{equation}
  \int \diff{x} P_1(x) = N \mdot
\end{equation}
Hence, in the end we recover exactly the same result as the one obtained by our method with the density (with the extra terms discussed in the previous section) and in \cite{pruessnerFieldTheoriesActive2022}, equation (S-V.109), with $\Upsilon=0$. The external potential can be easily added, we skipped it in the sake of conciseness.

\end{appendices}

\printbibliography

\end{document}